\documentstyle[psfig]{mn}

\newif\ifAMStwofonts

\ifoldfss
  \ifCUPmtlplainloaded \else
    \NewTextAlphabet{textbfit} {cmbxti10} {}
    \NewTextAlphabet{textbfss} {cmssbx10} {}
    \NewMathAlphabet{mathbfit} {cmbxti10} {} 
    \NewMathAlphabet{mathbfss} {cmssbx10} {} 
  \fi
  \ifAMStwofonts
    \ifCUPmtlplainloaded \else
      \NewSymbolFont{upmath} {eurm10}
      \NewSymbolFont{AMSa} {msam10}
      \NewMathSymbol{\upi}     {0}{upmath}{19}
      \NewMathSymbol{\umu}     {0}{upmath}{16}
      \NewMathSymbol{\upartial}{0}{upmath}{40}
      \NewMathSymbol{\leqslant}{3}{AMSa}{36}
      \NewMathSymbol{\geqslant}{3}{AMSa}{3E}

      \let\leq=\leqslant 
      \let\geq=\geqslant 
    \fi
  \fi
\fi 

\ifnfssone
  \newmathalphabet{\mathit}
  \addtoversion{normal}{\mathit}{cmr}{m}{it}
  \addtoversion{bold}{\mathit}{cmr}{bx}{it}
  \newmathalphabet{\mathbfit} 
  \addtoversion{normal}{\mathbfit}{cmr}{bx}{it}
  \addtoversion{bold}{\mathbfit}{cmr}{bx}{it}
  \newmathalphabet{\mathbfss} 
  \addtoversion{normal}{\mathbfss}{cmss}{bx}{n}
  \addtoversion{bold}{\mathbfss}{cmss}{bx}{n}
  \ifAMStwofonts
    \ifCUPmtlplainloaded \else
      \UseAMStwoboldmath
      \makeatletter
      \new@mathgroup\upmath@group
      \define@mathgroup\mv@normal\upmath@group{eur}{m}{n}
      \define@mathgroup\mv@bold\upmath@group{eur}{b}{n}
      \edef\UPM{\hexnumber\upmath@group}
      \new@mathgroup\amsa@group
      \define@mathgroup\mv@normal\amsa@group{msa}{m}{n}
      \define@mathgroup\mv@bold\amsa@group{msa}{m}{n}
      \edef\AMSa{\hexnumber\amsa@group}
      \makeatother
      \mathchardef\upi="0\UPM19
      \mathchardef\umu="0\UPM16
      \mathchardef\upartial="0\UPM40
      \mathchardef\leqslant="3\AMSa36
      \mathchardef\geqslant="3\AMSa3E

      \let\leq=\leqslant 
      \let\geq=\geqslant 
    \fi
  \fi
\fi 

\ifnfsstwo
  \DeclareMathAlphabet{\mathbfit}{OT1}{cmr}{bx}{it}
  \SetMathAlphabet\mathbfit{bold}{OT1}{cmr}{bx}{it}
  \DeclareMathAlphabet{\mathbfss}{OT1}{cmss}{bx}{n}
  \SetMathAlphabet\mathbfss{bold}{OT1}{cmss}{bx}{n}
  \ifAMStwofonts
    \ifCUPmtlplainloaded \else
      \DeclareSymbolFont{UPM}{U}{eur}{m}{n}
      \SetSymbolFont{UPM}{bold}{U}{eur}{b}{n}
      \DeclareSymbolFont{AMSa}{U}{msa}{m}{n}
      \DeclareMathSymbol{\upi}{0}{UPM}{"19}
      \DeclareMathSymbol{\umu}{0}{UPM}{"16}
      \DeclareMathSymbol{\upartial}{0}{UPM}{"40}
      \DeclareMathSymbol{\leqslant}{3}{AMSa}{"36}
      \DeclareMathSymbol{\geqslant}{3}{AMSa}{"3E}

      \let\leq=\leqslant 
      \let\geq=\geqslant 
    \fi
  \fi
\fi 

\ifCUPmtlplainloaded \else
  \ifAMStwofonts \else 
    \def\upi{\pi}
    \def\umu{\mu}
    \def\upartial{\partial}
  \fi
\fi

\title{The massive stellar content in NGC 604 and its evolutionary state.}
\author[Rosa M. Gonz\'alez Delgado \& Enrique P\'erez]
       {Rosa M. Gonz\'alez Delgado \& Enrique P\'erez\\
        Instituto de Astrof\'\i sica de Andaluc\'\i a (CSIC). Apdo. 3004, 18080 Granada, Spain}
\date{Accepted 1999.
      Received 1999;
      in original form 1999}

\pagerange{\pageref{firstpage}--\pageref{lastpage}}
\pubyear{1994}

\begin{document}

\maketitle

\label{firstpage}

\begin{abstract}

This paper analyzes the integrated ultraviolet spectra taken with the
{\it International Ultraviolet Explorer} ({\it IUE}) and optical ground-based
spectra taken with the William Herschel Telescope (WHT) of the giant H\,{\sc
ii} region NGC 604. These data are complemented with ultraviolet (WFPC2
through F170W) and H$\alpha$ (WFPC2 through F656N) images retrieved from the
{\it Hubble Space Telescope} ({\it HST}) archive.

The ultraviolet resonance wind stellar lines, the nebular optical emission
lines and the higher order terms of the Balmer series and HeI absorption lines
detected in the spectra of NGC 604 are interpreted using evolutionary models
optimized for young star forming regions. The evolutionary state and the
massive stellar content of the region is derived in a self-consistent way. 

The three techniques applied suggest that the central ionizing cluster in NGC
604 is very young, $\simeq3$ Myr old, and that the stars in the cluster were formed
in an instantaneous burst following a Salpeter ($\alpha$=2.35) or flatter
($\alpha$=1.5) initial mass function (IMF), having stars more massive that 80
M$\odot$. The stellar cluster is able to provide most of the ionizing
photons needed to photoionize the whole nebula, and the wind power to form the
central shell structure where the cluster core is located. The stellar
cluster is affected by an extinction similar to the average extinction 
that affects the ionized gas.
The estimated number of massive stars in the cluster is also in agreement with 
that derived from previous studies based on the detection of individual stars. 
The results that we present here support the use of these techniques for the 
interpretation of the integrated light of more distant star forming regions. 

\end{abstract}


\section{Introduction}

Giant extragalactic H\,{\sc ii} regions (RH\,{\sc ii}s) are amongst the
brightest objects in galaxies. RH\,{\sc ii}s have been studied extensively
because they are the best indicators of the conditions that lead to massive
star formation, and they show the cloud properties immediately after the moment
when stars form. Equally important, they are favourable places where to
measure the chemical abundances and star formation rates in galaxies, and
hence can be used to trace the chemical and star formation history of the
universe, when observed in galaxies at different redshift (Madau et al. 1996;
Kobulnicky \& Zaritsky 1999).

RH\,{\sc ii}s are characterized to have a size larger than 100 pc and H$\alpha$
luminosity brighter than 10$^{39}$ erg s$^{-1}$ (Kennicutt 1984). Therefore, 
the nebula requires an ionizing photon luminosity larger than 10$^{51}$ 
s$^{-1}$; this is provided by a stellar cluster that contains more than
100 young massive stars. These characteristics are very similar to those of
starburst galaxies. However, they are less luminous than prototypical
starbursts, and thus they are referred to as mini-starbursts (Walborn 1991). 
Like starbursts, RH\,{\sc ii}s show a nebular emission-line spectrum at optical
wavelengths, and an absorption-line spectrum at wavelengths shorter than the
Balmer jump (Leitherer 1997; Rosa, Joubert \& Benvenuti 1984). This spectral 
morphology reflects the fact that RH\,{\sc ii}s are powered by massive stars.
These stars emit photons with energies of tens of eV which are absorbed and
re-emitted in their stellar winds, producing ultraviolet resonance transitions. 
However, the stellar wind is optically thin to most of the ultraviolet  photons, 
that can travel tens of parsec from the star before they are absorbed and 
photoionize the surrounding  interstellar medium. Subsequently, this ionized
gas cools down via an emission spectrum. This spectral dichotomy picture
allows to derive the stellar content and the evolutionary state of the
cluster, through the analysis of the ultraviolet (e.g.  Vacca et al. 1995;
Leitherer et al. 1996), or the optical light (e.g. Cid Fernandes et al. 1992;
Garc\'\i a-Vargas \& D\'\i az 1994; Stasi\'nska \& Leitherer 1996) using 
evolutionary synthesis and photoionization models. The two techniques have 
been applied to the prototypical starburst nucleus NGC 7714, giving similar 
results (Gonz\'alez Delgado et al. 1999a).  

However, around the Balmer jump the spectra of starbursts 
(Gonz\'alez Delgado et al. 1998) and some H\,{\sc ii} regions 
(Terlevich et al. 1996) show the higher order terms of the Balmer series 
and He\,{\sc i} lines in absorption, formed in the photospheres of massive
stars. These stellar lines can be detected in absorption because the strength
of the gaseous Balmer lines in emission decreases rapidly with decreasing
wavelength, whereas the equivalent width of the stellar absorption lines is
almost constant with wavelength (Gonz\'alez Delgado, Leitherer \& Heckman 1999b). 
Evolutionary synthesis models that predict the profiles of the higher order 
terms of the Balmer series and He\,{\sc i} lines in absorption can also be 
used to estimate the evolutionary state of the stellar cluster in H\,{\sc ii} 
regions.

The second more luminous and brightest H\,{\sc ii} region in the Local Group of
galaxies, after 30 Dor in the LMC, is NGC 604 in M 33. Its distance, 840 kpc  
(Freedman et al. 1991), allows detailed studies of the individual stars, 
and also to obtain integrated properties. Thus, NGC 604 is an excellent 
laboratory in which to explore questions about the effect of star formation, 
the slope and upper mass limit of the IMF, and to test the consistency between 
the three techniques described above, that allow to determine the stellar content  
and the evolutionary state of the ionizing stellar cluster from the spatially 
integrated ultraviolet and optical spectra of the H\,{\sc ii} region. 

NGC 604 has been intensively studied in the past (e.g. Peimbert 1970; Israel \&
van der Kruit 1974; Smith 1975; Hawley \& Grandi 1977; Kwitter \& Aller 1981;
Conti \& Massey 1981; Rosa \& D'Odorico 1982; Viallefond \& Goss 1986). A
precise determination of the chemical composition of the gas was made
by D\'\i az et al. (1987); they give an oxygen abundance 12+log(O/H)=8.5.
VLA observations indicate that the gas is very tenuous (average electron
density, $rms$ $N_{\rm e}$, between a few and $\sim100$ cm$^{-3}$), and not very
dusty, with a mean visual extinction $A_V\sim0.5$ mag (Churchwell \&
Goss 1999). The morphology of the ionized gas is very complex, showing many
filaments and shell structures that are expanding (e.g. Hippelein \& Fried
1984; Rosa \& Solf 1984; Clayton 1988; Sabalisck et al. 1995;
Mu\~noz-Tu\~n\'on et al. 1995; Yang et al. 1995; Medina Tanco et al. 1997). 
This morphology is a consequence of the violent star formation activity in NGC 604.
{\it HST} imaging photometry has revealed that the stellar cluster is resolved into
$\sim$200 massive stars in an area of $\sim$10000 pc$^2$ (Drissen, Moffat \&
Shara 1993; Hunter et al. 1996). Evidences that the stellar cluster is evolved
come from the detection of Wolf-Rayet stars (Conti \& Massey 1981; D'Odorico
\& Rosa 1981; Drissen, Moffat \& Shara 1990, 1993), one supernova remnant
(D'Odorico et al. 1980), and one candidate to red supergiant (Terlevich et al. 1996). 
The existence of these stars as members of the stellar cluster of NGC 604 
suggests that the age of the region is in the range 3-5 Myr. 

This paper presents spatially integrated ultraviolet and optical spectra of NGC 604. 
The goal is to derive the massive stellar content and the evolutionary state of the
ionizing cluster, by means of evolutionary synthesis and photoionization
models applied to the ultraviolet resonance wind stellar lines, to the nebular
emission lines and to the higher-order terms of the Balmer series and He\,{\sc
i} lines in absorption. The consistency  between the results obtained with
the three techniques, and the similarity with the results obtained from
studies based on the detection of individual stars, strengthens the reliability
and power of these techniques when they are applied to determine the stellar
content of more distant star-forming regions. Section 2 presents the observations. 
Section 3 describes the photoionization models that fit the emission-line
spectrum of the nebula. In section 4, the higher-order terms of the Balmer
and He\,{\sc i} lines in absorption are analyzed. Evolutionary synthesis models
of the ultraviolet resonance wind stellar lines are in section 5. In section
6, the massive stellar content is derived. The summary and conclusions are in
section 7.

\section{Observations}

\subsection{{\it HST} images}

NGC 604 has been intensively observed with the WFPC2 camera onboard the {\it HST}. 
For the purpose of this work, we have retrieved {\it HST} archive images at the 
ultraviolet (F170W filter) and optical H$\alpha$ (filter F656N) wavelengths.
Two F170W exposures of 350 s each and two F656N exposures of 1000 s each  were
combined to produce one final ultraviolet (Fig. 1) and one H$\alpha$ image
(Fig. 2). The nebula was centred in the PC camera, therefore the
ionizing cluster is sampled at 0.046 arcsec~pixel$^{-1}$, corresponding to
0.19 pc~pixel$^{-1}$ (at a distance of 840 kpc, 1 arcsec corresponds to 4.1
pc). The ultraviolet emission is spread in the inner 20$\times$20 arcsec. The
stellar density looks much lower than that observed in super star clusters.
Surface brightness photometry indicates that 75 percent of the ultraviolet flux 
is produced by the core of the stellar cluster, that is contained within a 
central nebular hole of diameter $\sim15$ arcsec; we shall refer to this as the 
centre of the nebula. The H$\alpha$ image shows many filaments and shell structures 
that extend out to $\sim200$ pc from the centre of the nebula. However, 
the brightest parts of the nebula trace almost a ring structure at a distance of 
40 pc from the centre.

\setcounter{figure}{0}

\begin{figure}
\caption{{\it HST\/}+WFPC2 (F170W) image of NGC 604 at ultraviolet
wavelengths.  The scale is 0.046 arcsec pixel$^{-1}$, equivalent to 0.19 pc
pixel$^{-1}$.  The rectangles mark the positions of the {\it IUE} apertures.
North is up and East to the left.} 
\label{HSTuv} 
\end{figure}

\begin{figure}
\caption{{\it HST\/}+WFPC2 (F653N) image in the H$\alpha$ emission line (grayscale). 
The ultraviolet emission (F170W) is overlayed in white contours. The scale is 0.1
arcsec pixel$^{-1}$.   The lines indicate the slits of the ground-based optical
spectroscopy. Note that the core of the cluster is within a nebular hole of 20
pc radius. North is up and East to the left.} 
\label{HSTha}
\end{figure}

The total ultraviolet flux per unit wavelength in the F170W image is
$1.9\times10^{-13}$  erg s$^{-1}$ cm$^{-2}$ \AA$^{-1}$. The H$\alpha$ flux is
$4.0\times10^{-11}$ erg s$^{-1}$ cm$^{-2}$, in good agreement with the
value measured by Churchwell \& Goss (1999). 

We perform surface brightness photometry in the H$\alpha$ image using a 
circular aperture around the central cluster nebular hole. 
The cumulative flux is plotted in Fig. 3. This plot shows that 50, 75 
and 90 percent of the total flux is contained in the inner 60, 100 and 140 pc 
radii\footnote{Unless otherwise specified, distances refer to radial distances
from the adopted centre of symmetry, as shown in Figs. 1 and 2.},
respectively. The H$\alpha$ flux distribution of NGC 604 has the characteristic
features of a core-halo distribution, where the core can be quantitative and
qualitatively explained by a thick shell structure. A detailed modelling of the
geometry of NGC 604 will be presented elsewhere. The surface brightness weighted core
radius  of the region (following the method described in Garc\'\i a-Vargas et al.
1997)  is 80 pc. 

The $rms$ electron density is also estimated from the H$\alpha$ image as 
$(Q/\alpha_BV)^{1/2}$, where $Q$ is the ionizing photon luminosity,
$\alpha_B$ the total recombination coefficient of H, and $V$ the volume of the
ionized region. Fig. 3 also shows the $rms~N_{\rm e}$ as a function of
distance, that ranges from values between less than 100 cm$^{-3}$ to a few
cm$^{-3}$, with $rms~N_{\rm e}\sim10$ cm$^{-3}$ at 100 pc from the center of the
nebula.

\begin{figure}
\psfig{figure=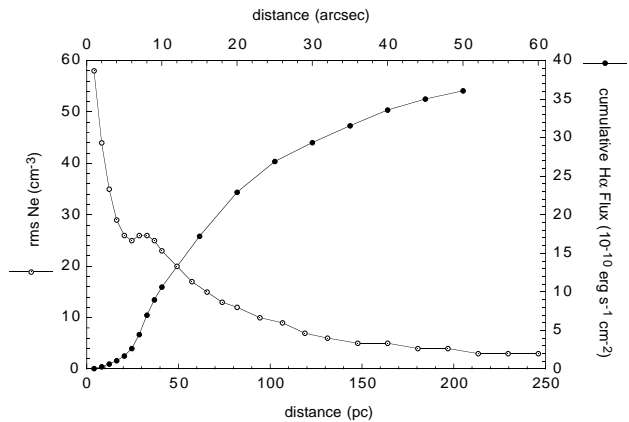,width=85mm,angle=0}
 \caption{Cumulative H$\alpha$ flux (full dots) and the $rms$ electron
density (open dots) measured in the WFPC2 (F653N) image, using a circular
aperture as a function of the distance to the center of the nebula.}
\label{Hacum} 
\end{figure}

\subsection{Ultraviolet spectroscopy}

NGC 604 has been observed with the {\it IUE} through the large aperture 
(9.5$\times$22 arcsec)\footnote{The precise form and size of the {\it IUE} 
large aperture is not known; we take a size of $22\times9.5$ arcsec. 
The actual form is somewhat intermediate between the rectangle shown here 
and an enclosed ellipse.} in low-dispersion mode, with either the SWP 
(1100--1900 \AA) or the LWR (1900--3200 \AA) cameras. The spectra were taken
with  the large aperture located at several different positions across the
region.  We have retrieved from the {\it IUE} Newly Extracted Spectra (INES,
Rodr\'\i guez-Pascual et al. 1999; Cassatella et al. 2000) archive the spectra 
taken at the positions shown in Fig. 1 (see also Table \ref{tab1}). Several 
of the spectra were also retrieved in February 1997 from the {\it IUE} ULDA 
archive. When comparing the two different extractions  of the same aperture 
spectrum, we noticed that the slope of the spectra  were slightly steeper in 
the ULDA archive; both extractions have the same flux at 1500 \AA\  but the 
ULDA extraction is 10 percent brighter at 1280 \AA\ than the INES extraction.  
ULDA used a different extraction algorithm ({\sevensize IUESIPS}) than INES, 
and their discrepancies are well documented (Gonz\'alez-Riestra et al. 2000).
INES represents the state of the art in the extraction of {\it IUE} spectra; 
so this is the data set that we use.

The flux at 1750 \AA\ measured in the spectra SWP5688 (PA=149$\degr$) and SWP6638 
(PA=110$\degr$) is $1.6\times10^{-13}$ erg s$^{-1}$ cm$^{-2}$ \AA$^{-1}$
and $1.3\times10^{-13}$ erg s$^{-1}$ cm$^{-2}$ \AA$^{-1}$, respectively. 
The sum total flux in these two apertures represents approximately the total 
ultraviolet flux emitted by the stellar cluster in NGC 604, and it amounts to 
$2.9\times10^{-13}$ erg s$^{-1}$ cm$^{-2}$ \AA$^{-1}$.
This value is 45 percent larger than the total monochromatic flux measured 
in the F170W {\it HST} image 
($\rm F_{170}=1.9\times10^{-13} erg\ s^{-1}\ cm^{-2} \AA^{-1}$). 
Fig. 4 shows the {\it IUE} spectra at PA=110$\degr$ and at PA=32$\degr$,
together with the average spectrum of those at PA=149$\degr$, 136$\degr$
and 159$\degr$.

\begin{figure}
\psfig{figure=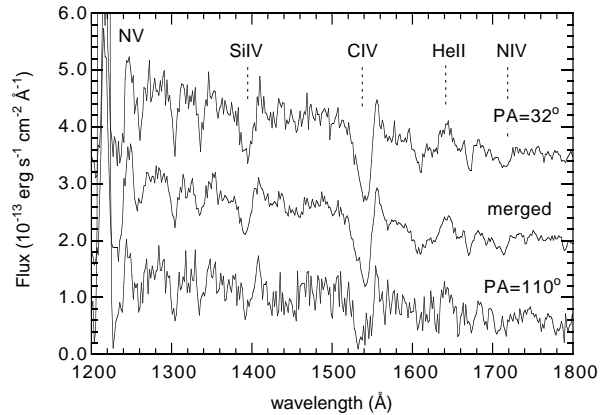,width=85mm,angle=0}
\caption{INES spectra of NGC 604 taken at PA=32$\degr$, and at
PA=110$\degr$,  and the average spectrum of those at PA=136$\degr$, 149$\degr$ and
159$\degr$.  The spectra are dereddened by $E(B-V)=0.03$ using the MW Galactic
extinction law.  For the sake of clarity, the spectra at PA=32$\degr$ and
110$\degr$ are plotted  after adding an artificial bias of $\pm1.5\times10^{-13}$
erg s$^{-1}$ cm$^{-2}$ \AA$^{-1}$,  respectively. The most important wind resonance
ultraviolet stellar lines are labelled. Other important absorption lines visible
are formed in the intervening interstellar medium, and are of no concern to us
here.}  
\label{IUEspec} 
\end{figure}

\setcounter{table}{0}
\begin{table*}
\centering
\begin{minipage}{140mm}
\caption{Journal of Observations.}
\label{tab1}
\begin{tabular}{@{}ccccrrl@{}}
Telescope & Instrument & Filter or Grating & wavelength & P.A. & t$\rm_{exp}~(s)$ &
Image Name   \\
\\
{\it HST} & WFPC2 & F170W & $\lambda_c$=1750 \AA, $\Delta\lambda$=545 \AA& -- & 2$\times$350 & U2C60B01T, U2C60B02T \\  
{\it HST} & WFPC2 & F656N & $\lambda_c$=6564 \AA, $\Delta\lambda$=21 \AA & -- & 2200         & U2LX0301A            \\
\\
{\it IUE} & SWP   &       & 1100--1900 \AA &  32$\degr$ &  12000         & SWP19449  \\
{\it IUE} & SWP   &       & 1100--1900 \AA & 136$\degr$ &  15300         & SWP19181  \\
{\it IUE} & SWP   &       & 1100--1900 \AA & 159$\degr$ &   6330         & SWP7349   \\
{\it IUE} & SWP   &       & 1100--1900 \AA & 149$\degr$ &   6000         & SWP5682   \\
{\it IUE} & SWP   &       & 1100--1900 \AA & 110$\degr$ &   4800         & SWP6638   \\
\\
WHT & ISIS  & G300B & 3700-5500 \AA & 131$\degr$ &  2$\times$3600 &        \\
WHT & ISIS  & G300B & 3700-5500 \AA &  60$\degr$ &  3600          &        \\
WHT & ISIS  & G300B & 3700-5500 \AA & 120$\degr$ &  1800          &        \\
WHT & ISIS  & G316R & 7900-9700 \AA &  60$\degr$ &  3600          &        \\
WHT & ISIS  & G316R & 7900-9700 \AA & 120$\degr$ &  1800          &        \\
\end{tabular}
\end{minipage}
\end{table*}

\subsection{Optical spectroscopy}

Optical spectra were obtained with the 4.2m William Herschel Telescope at the
Observatorio del Roque de los Muchachos (La Palma), as part of the
GEFE\footnote{Grupo de Estudios de Formaci\'on Estelar, is an international
collaboration formed to take advantage of the international time granted by
the Comit\'e Cient\'\i fico Internacional at the Observatories in the Canary
Islands.} collaboration. The details of the observations and data reduction
are in Terlevich et al. (1996). For the purpose of this work, we use the
scanned spectra at PA=60$\degr$ and at PA=120$\degr$, and the single
long-slit spectrum at PA=131$\degr$. The scanned spectra were centred at
R.A.=1$\rm^h$~31$\rm^m$~43$\rm^s$ and  Dec=30$\degr$~31{\arcmin}~52{\arcsec},
and they cover the core of the region by displacing a 1 arcsec wide longslit 
in steps of 1 arcsec and taking at each position a 1 minute exposure.  
The process is repeated until an area of
1.75{\arcmin}$\times$0.18{\arcmin}  (for the scan at PA=120$\degr$) and
1.75{\arcmin}$\times$0.23{\arcmin}  (for the scan at PA=60$\degr$) are covered
(see Fig. 2). A one dimensional  spectrum representative of the inner region
of the nebula was obtained by merging the  extractions that cover
11{\arcsec}$\times$60{\arcsec} (at PA=120$\degr$) and 
14{\arcsec$\times$60{\arcsec} (at PA=60$\degr$). At PA=131$\degr$, a
one-dimensional  spectrum was extracted of length 4.6 arcsec, and centered at
the position where the continuum  is maximum (see fig. 3b in Terlevich et
al. 1996). At this position the fraction of the stellar continuum with respect
to the nebular emission lines is maximum, because it corresponds to the zone 
of the nebular hole. As previously noted by Terlevich et al. (1996), the
spectrum shows the  higher-order terms of the Balmer series and He\,{\sc i}
lines in absorption, and the continuum  represents well the stellar light of
the ionizing stellar cluster in NGC 604 (Fig. 5).

\begin{figure}
\psfig{figure=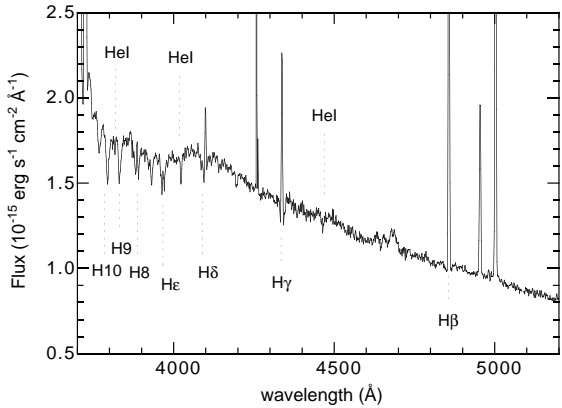,width=85mm,angle=0}
\caption{One-dimensional ground-based optical spectrum of NGC 604 obtained 
through a $1.2\times4.6$ arcsec aperture at the position where the continuum 
is maximum (see fig. 3b in Terlevich et al. 1996). The high-order terms of the 
Balmer series and He\,{\sc i} $\lambda$4471, He\,{\sc i} $\lambda$4026 and 
$\lambda$3819 are labeled.}  
\label{optspec} 
\end{figure}

\section{Modelling the nebular emission-lines}

The emission-line spectrum of an RH\,{\sc ii} depends on the radiation field 
from the ionizing stellar cluster, and on the density distribution and chemical 
composition of the gas. A photoionization code takes as input the spectral 
energy distribution of the cluster, and it solves the ionization-recombination 
and heating-cooling balances, to predict the ionization structure of the nebula, 
the electron density and the intensity of the emission lines. By comparing 
these output with observations, it is possible to obtain information about 
the ionizing stars and their evolutionary state. 

The star formation law, age and massive stellar content of the stellar 
clusters can be constrained by comparing the observed emission line strengths 
with the predictions from the photoionization models, when the code uses as 
input the spectral energy distribution (SED) generated by a stellar evolutionary 
synthesis code. This technique has been used successfully to study the stellar 
content in starbursts and RH\,{\sc ii}s (e.g. Garc\'\i a-Vargas, Bressan 
\& D\'\i az 1995a,b; Garc\'\i a-Vargas et al. 1997; Gonz\'alez Delgado et al.
1999a; Luridiana, Peimbert \& Leitherer 1999; Stasi\'nska \& Schaerer 1999). Here,
we fit the emission-line spectrum of NGC 604 using the photoionization code
{\sevensize CLOUDY}  (version 90.04, Ferland 1997).

\subsection{Input parameters}

In order to predict the intensities of the emission lines, we fix the geometry 
of the nebula, the electron density and the chemical composition of the gas. 
We assume that the constant density gas is ionization bounded and spherically 
distributed around the ionizing cluster. The inner radius is 20 pc 
(as seen in the H$\alpha$ image), and the outer radius is determined by 
the ionization front. The results do not depend strongly on the inner 
radius if this changes by a factor 2 (e.g. Luridiana, Peimbert \& Leitherer 1999). 
We assume that the gas  occupies only a fraction of the sphere. Thus the 
filling factor, $\phi$, is a free parameter  with values of 10$^{-1}$,  
$5\times10^{-2}$, 10$^{-2}$, 10$^{-3}$ and 10$^{-4}$. The change of filling 
factor is equivalent to changing the ionization parameter $U$, defined as
$Q/(4\pi R N_{\rm e} c$); where $Q$ is the ionizing photon luminosity, 
$N_{\rm e}$ the electron density, $c$ the speed of light and $R$ the distance
of the gas to the ionizing source. For a spherical geometry, the average $U$ 
is proportional to $(\phi^2N_{\rm e}Q)^{1/3}$.

The chemical abundances are fixed to those derived by D\'\i az et al. (1987). 
We scale the chemical composition of the gas to the value of the oxygen abundance, 
12+log(O/H)=8.5, except the abundance of N, Ne, S and He, for which we take 
the values derived by D\'\i az et al. (1987; see Table 2). The electron density 
derived by D\'\i az et al. (1987) is of the order of 100 cm$^{-3}$. Our 
measurements of the [S\,{\sc ii}]6717/6731 ratio indicate the low-density limit,
with values below 100 cm$^{-3}$ \footnote{Notice, however, that the atomic 
parameters for the sulfur have changed in the intervening time}.  We have computed 
constant density models at two values of the electron density, 100 cm$^{-3}$ and
30 cm$^{-3}$.  The latter value comes from the filling factor and an $rms$ electron
density of 10 cm$^{-3}$, derived from the H$\alpha$ image. Also, an electron
density of 30 cm$^{-3}$ predicts a [S\,{\sc ii}]6717/6731 ratio which is  in
better agreement with the observed value.

\begin{table}
\centering
\begin{minipage}{40mm}
\caption{Input parameters to {\sevensize CLOUDY}.}
\label{tab2}
\begin{tabular}{@{}ll@{}}
             &         \\
12+log(O/H)  &  8.51   \\
log (He/H)   & -1.08   \\
log (N/H)    & -4.65   \\
log (Ne/H)   & -4.27   \\
log (S/H)    & -5.05   \\
log Q        & 51.54 (s$^{-1}$) \\
Inner Radius & 20 pc   \\
Ne           & 30, 100 cm$^{-3}$ \\
\end{tabular}
\end{minipage}
\end{table}

\subsection{The radiation field}

The radiation field used as input to {\sevensize CLOUDY} is the spectral energy 
distribution generated by the evolutionary synthesis code developed by Leitherer 
and colaborators (Leitherer et al. 1999). The code includes the new set of stellar 
evolutionary models of the Geneva group (Schaller et al. 1992; Schaerer et al. 
1993a,b; Charbonnel et al. 1993; Meynet et al. 1994), and the stellar atmospheres 
grid compiled by Lejeune et al. (1997), supplemented by the expanding spherical 
non-LTE models of Schmutz, Leitherer \& Gruenwald (1992). The latter stellar 
atmosphere models are applied to stars with very strong stellar winds. The 
spectral energy distribution was generated using the Z=0.008 (Z$\odot$/2.5) 
metallicity tracks, assuming that stars and gas have the same metallicity. 
We assume two different star formation scenarios: instantaneous burst and 
continuous star formation at a constant rate. Different models are also computed 
with different assumptions about the IMF. The slope is Salpeter ($\alpha$= 2.35), 
flatter ($\alpha$= 1.5), or steeper ($\alpha$= 3.3). The upper mass limit 
cut-off is set to 30, 60, 80 or 120 M$\odot$. The lower mass limit is fixed 
to 1 M$\odot$. However, photoionization models are not influenced by the lower 
limit mass cut-off if it is below 10 M$\odot$. The spectral energy distribution 
is normalized to an ionizing  photon luminosity of $\rm log\,Q=51.54$ s$^{-1}$. 
This value is derived from the total H$\alpha$ flux measured in the image, 
after correcting by the mean extinction, $A_V=0.5$ mag, derived by Churchwell 
\& Goss (1999). We will check below the compatibility of this normalization 
with the mass of the stellar cluster derived from the ultraviolet continuum 
luminosity.

\subsection{The observational constraints}

The emission lines were measured in our scanned spectrum of the nebula. 
The observed ratios have been corrected by the mean extinction, c(H$\beta$)=0.22, 
derived by Churchwell \& Goss (1999). This value is in agreement with the 
reddening derived from the H$\gamma$/H$\beta$ ratio measured in the scanned 
spectrum, when this ratio is corrected by the underlying stellar absorption. 
This extinction is lower than the values derived by D\'\i az et al. (1987), 
c(H$\beta$)=0.3-0.4; however, the extinction changes across NGC 604 
(Ma\'\i z-Apell\'aniz 1999). The de-reddened emission line ratios to be fitted 
by the photoionization models are included in Table \ref{tab3}. This table also 
includes the emission line ratios derived by D\'\i az et al. (1987) in several 
parts of the nebula, and those from an integrated spectrum in Ma\'\i z-Apell\'aniz 
(1999). The range of values for each of these emission line ratios defines the
tolerance that we accept for the difference between the observed and the predicted
values from the models.

Following Stasi\'nska \& Schaerer (1999), we try to fit the strength 
of the emission lines, and also emission-line ratios indicative of the
electron temperature ([O\,{\sc iii}]4363/5007), electron density 
([S\,{\sc ii}]6717/6732)  and ionization structure 
([O\,{\sc iii}]5007/[O\,{\sc ii}]3727,  
[S\,{\sc iii}]9069/[S\,{\sc ii}]6716+6732).
Another important observational constraint is the  radius of the nebula. The
H$\alpha$ image shows that the flux extends out to 200 pc  of the center of
the region; however, more than 90, 75 and 50 percent of the total  flux is
within the inner 140 pc, 100 pc and 60 pc radii, respectively (Fig. 3).  Our
determination of the core radius of the region is 80 pc.

\subsection{Model results}

We start by fixing the filling factor and the electron density by means of 
fitting the line ratios [S\,{\sc ii}]6717+6731/H$\beta$ and [S\,{\sc ii}]6717/6731, 
and the radius of the nebula. [S\,{\sc ii}]6717+6731/H$\beta$ is a good calibrator 
of the ionization parameter, because this ratio does not depend much on the IMF
assumptions, on the star formation law or on the evolutionary state of the
cluster (Gonz\'alez Delgado et al. 1999a). Instead, this ratio depends on the
filling factor, and for fixed geometry, ionizing photon luminosity and
electron density, it depends on the ionization parameter\footnote{Nonetheless,
we should always be cautious when using sulfur dependent diagnostics, given the
uncertainties about the atomic parameters and its not so well known ionization;
also this calibration relies on the assumption of an ionization bounded nebula.}. 
The observed ratio, [S\,{\sc ii}]6717+6731/H$\beta\simeq0.4$, indicates a filling 
factor $\phi\simeq0.1$ (Fig. 6). Therefore, a large fraction of the volume
of the region is filled with ionized gas; however, this gas is very tenuous
because the $rms$ electron density measured in the H$\alpha$ image is
$\simeq10$ cm$^{-3}$, and the electron density in the S$^{+}$ zone, derived
from the definition of filling factor  ($\phi=(N_{\rm e}(rms)/Ne)^2$), is
$N_{\rm e}\simeq30$ cm$^{-3}$. The observed ratio [S\,{\sc ii}]6717/6731=1.4,
indicates that the electron density is $\leq100$ cm$^{-3}$. Models with a
filling factor of 0.1 and $N_{\rm e}=30$ cm$^{-3}$ predict well the observed 
[S\,{\sc ii}]6717/6731 ratio. If the electron density were as high as 
$N_{\rm e}=100$ cm$^{-3}$, then the filling factor would have to be lower, 
$\phi$=0.01, and in this  case the values of [S\,{\sc ii}]6717+6731/H$\beta$ 
and [S\,{\sc ii}]6717/6731 predicted  by the models are 0.9 and 1.3, respectively. 
These predictions are larger and smaller than the observed values, respectively. 
Furthermore, models with $\phi=0.1$ and $N_{\rm e}=30$ cm$^{-3}$ predict a 
Str\"omgren 
radius $\simeq$110 pc, in agreement with the radius derived from the H$\alpha$ 
surface brightness photometry. Models with $N_{\rm e}=100$ cm$^{-3}$ predict 
a radius a factor two smaller than models with $N_{\rm e}=30$ cm$^{-3}$. 
Therefore, we conclude that a filling factor $\phi\simeq0.1$ and a density 
$N_{\rm e}=30$ cm$^{-3}$ fit well the ratios [S\,{\sc ii}]6717+6731/H$\beta$ 
and [S\,{\sc ii}]6717/6731, and the size of the region. 

\begin{figure}
\psfig{figure=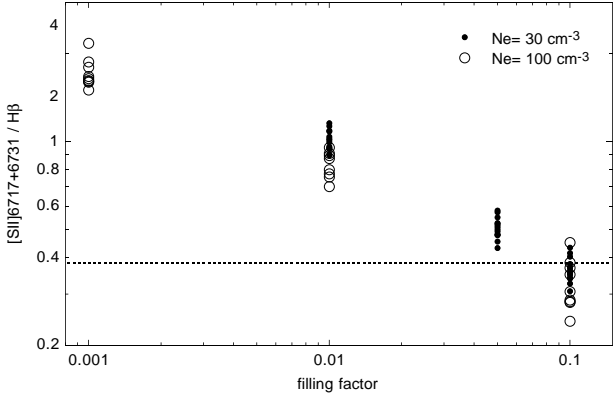,width=85mm,angle=0}
\caption{{\sevensize CLOUDY} predictions of the emission-line ratio 
[S\,{\sc ii}]6717+6732/H$\beta$ as a function of the filling factor, 
taking as input the spectral energy distributions from the evolution 
of a cluster formed in an instantaneous burst. Ratios obtained with 
electron density of 100 cm$^{-3}$ and 30 cm$^{-3}$ are plotted as open 
and filled circles, respectively. The observed value is plotted as 
a horizontal line.} 
\label{ffs2} 
\end{figure}

The emission-line ratios [O\,{\sc iii}]5007/H$\beta$, [O\,{\sc ii}]3727/H$\beta$, 
[O\,{\sc i}]6300/H$\beta$, and [N\,{\sc ii}]6584/H$\beta$ have been plotted as 
a function of the cluster age, using a filling factor $\phi$= 0.1 and an 
electron density $N_{\rm e}=30$ cm$^{-3}$. To check whether the models also fit 
line ratios indicative of the structure of the region, we have plotted the 
$\eta$ parameter, defined as 
([O\,{\sc iii}]5007+4959/[O\,{\sc ii}]3727)/([S\,{\sc iii}]9069+9532]/[S\,{\sc ii}]6717+6731), 
and the electron temperature predicted by the models as a function of the age.
The parameter $\eta$ is a measure of the softness of the radiation field 
(V\'\i lchez \& Pagel 1988). The models assume two different star formation
laws, an instantaneous burst or  continuous star formation at a constant rate,
and different assumptions about the  slope ($\alpha$=2.35, 1.50, 3.0) and upper
mass limit cut-off of the IMF ($M_{\rm up}=$120, 80, 60, 30 M$\odot$).
  
Fig. 7 compares the observed emission line ratios with the prediction of
continuous and instantaneous burst, for a Salperter IMF and upper mass limit
cut-off of 80 M$\odot$. The observed ratios are in better agreement with burst
than with continuous  star formation models (that predict high excitation
lines and [O\,{\sc i}]6300/H$\beta$ larger than observed). The continuous star
formation scenario can be made compatible with the observed emission-line
ratios if stars more massive than 50 M$\odot$ are not formed in the cluster;
however, these models predict significantly fewer numbers of Wolf-Rayet stars
in the cluster. Thus, continuous star formation models cannot fit the
emission-line ratios and  predict the existence of Wolf-Rayet stars
simultaneously.

Fig. 8 compares the observed emission line ratios with the prediction of an
instantaneous burst, for different assumptions of the IMF upper mass limit
cut-off. Models with $M_{\rm up}=30$ M$\odot$ predict high excitation lines
(low excitation lines) ratios  which are much lower (higher) than the observed
values. Thus, similar results  are obtained if $M_{\rm up}\leq50$ M$\odot$. Burst
models with $M_{\rm up}\geq60$ M$\odot$ fit the emission-line ratios if the age of
the cluster is 2.5--3 Myr or 4.5--4.8 Myr. However, if $M_{\rm up}=60$ M$\odot$, 
Wolf-Rayet stars only appear  in the cluster if it is 4--5 Myr old. 
Models with $M_{\rm up}=120$ M$\odot$ and 80 M$\odot$ give similar results
after the first 2 Myr, and the same behaviour with age.  The electron
temperature, $\eta$ and the emission-line ratios, all indicate that massive
stars ($M_{\rm up}\geq80$ M$\odot$) have to be present in the cluster, and that the
cluster is $\simeq3$ Myr or $\simeq4.5$ Myr old. Further constraints on the
upper mass limit cut-off will be imposed by the relavite number of Wolf-Rayet
with respect to O stars in the cluster. 

Burst models  have been computed also for different values of the IMF slope 
($\alpha$=2.35, 3.0 and 1.5).  The behaviour of the emission line ratios with 
the age is very similar for these three values of the IMF slope (Fig. 9). 
The observed ratios indicate an age $\simeq3$ Myr or 4.5 Myr, and cannot 
discriminate between the different values of the slope. Further constraints 
on the IMF slope will come from the WR/O ratio. This point will be discussed  
further in section 6.

\begin{figure*}
\psfig{figure=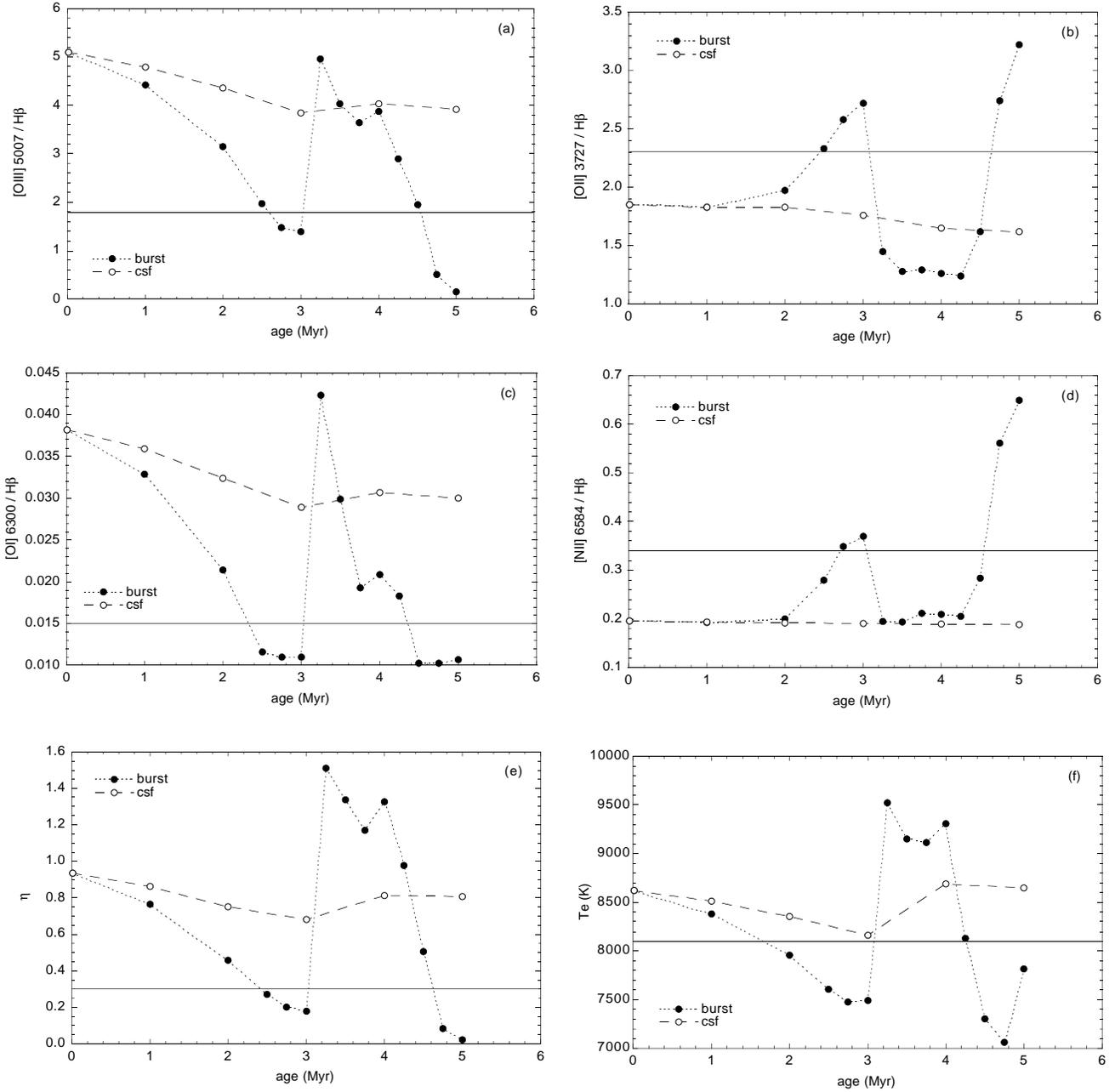,width=180mm,angle=0}
\caption{Predicted emission-line ratios, $\eta$ parameter and the electron 
temperature as a function of the cluster age, for an instantaneous burst 
(filled symbols) and continuous star formation (open symbols). The models 
assume that the mass of the cluster is distributed following a Salpeter IMF 
between 1 and 120 M$\odot$. The filling factor is 0.1 and the electron density 
30 cm$^{-3}$. The observed values are represented by an horizontal line.} 
\label{lineratios}
\end{figure*}

\begin{figure*}
\psfig{figure=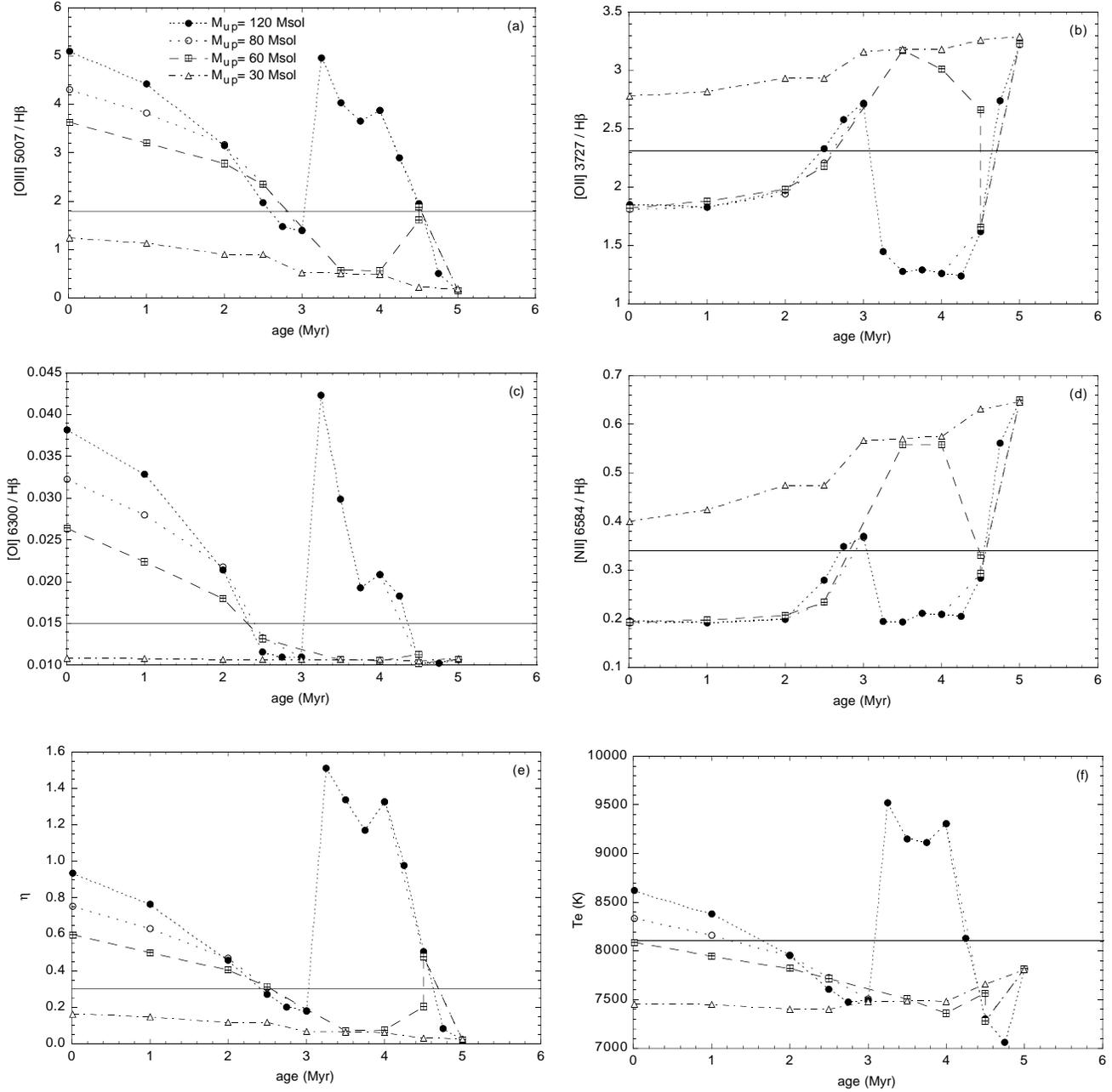,width=180mm,angle=0}
\caption{As Figure 7, but for an instantaneous burst with 
a Salpeter IMF upper mass limit cut-off 120 M$\odot$ (filled circles), 80 M$\odot$
(open circles), 60 M$\odot$ (squares), and 30 M$\odot$ (open triangles).} 
\label{Mupratios} 
\end{figure*}

\begin{figure*}
\psfig{figure=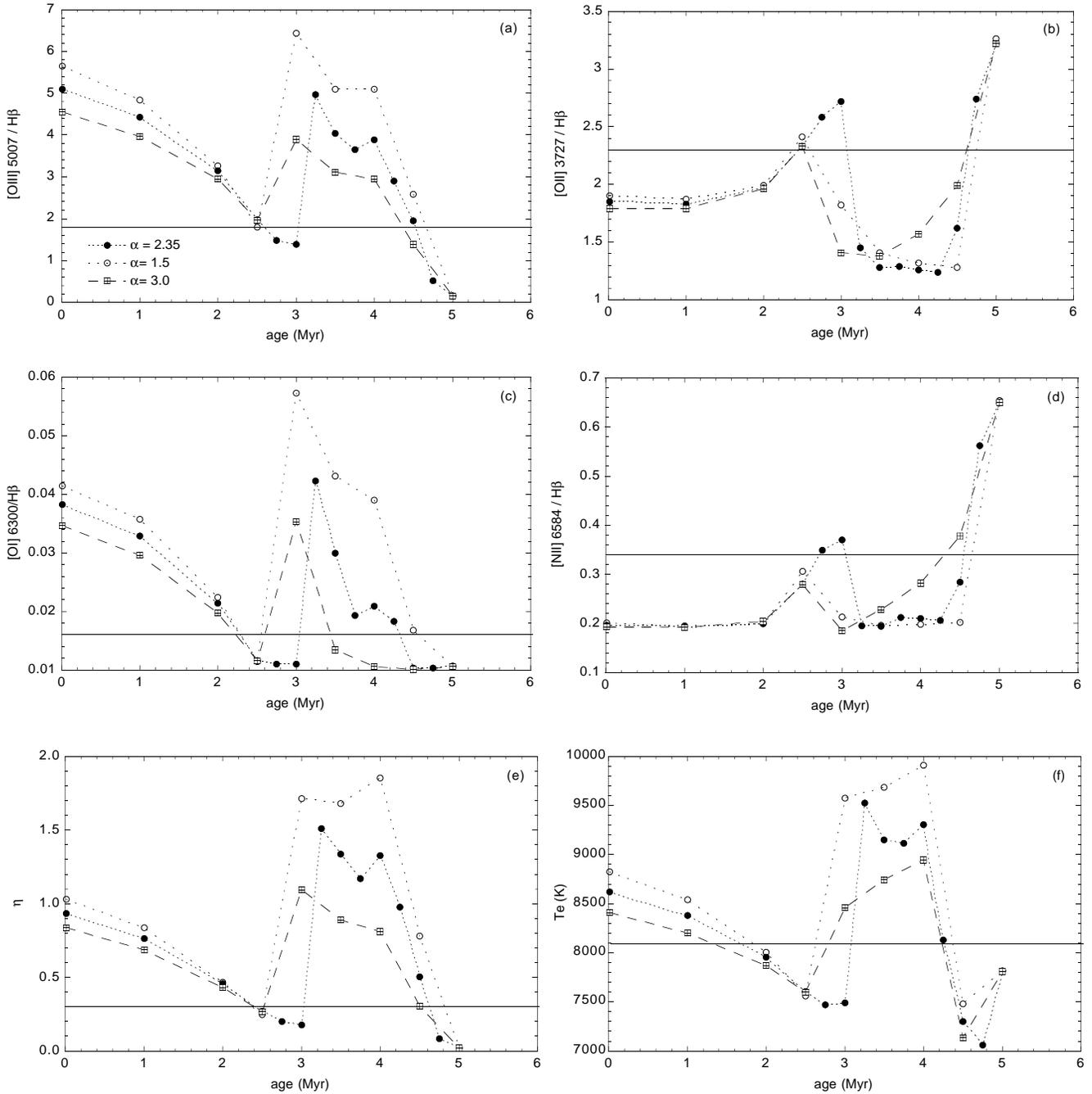,width=180mm,angle=0}
\caption{As Figure 7, but for an instantaneous burst with an IMF slope 
$\alpha$=2.35 (Salpeter, filled circles), $\alpha$= 1.5 (open circles) 
and $\alpha$=3.0 (squares) and upper mass limit cut-off 80 M$\odot$.} 
\label{sloperatios}
\end{figure*}

\begin{table*}
\centering
\begin{minipage}{150mm}
\caption{Emission line ratios, normalized to H$\beta$=1.}  
\label{tab3} 
\begin{tabular}{@{}llllll|lll@{}}
Line                    & \multicolumn{5}{c}{\underline{Observations}} & \multicolumn{3}{c}{\underline{Models}}  \\
                        & (1)    &   A    &  C          &   D   &   B    & 2.5 Myr & 3.0 Myr & 3.5 Myr \\
\hline
{[O\,{\sc ii}]} 3727            & 2.3    & 2.08   & 2.49        & 1.61  & 2.00   & 2.3     & 2.7     & 1.3     \\   
{[Ne\,{\sc iii}]} 3869          & 0.11   & 0.15   & 0.26        & 0.13  & 0.07   & 0.08    & 0.034   & 0.29    \\  
{[O\,{\sc iii}]} 4363           & 0.0054 & 0.0062 & $\leq0.004$ & 0.087 & 0.0063 & 0.0052  & 0.0033  & 0.024   \\  
{[O\,{\sc iii}]} 5007           & 1.8    & 2.69   & 0.97        & 2.55  & 1.8    & 2.0     & 1.4     & 4.0     \\ 
{[O\,{\sc i}]} 6300             &  --    & 0.015  &  --         & 0.01  & 0.02   & 0.012   & 0.011   & 0.03    \\ 
{[N\,{\sc ii}]} 6584            & 0.34   & 0.30   & 0.51        & 0.27  & 0.31   & 0.28    & 0.37    & 0.19    \\ 
{[S\,{\sc ii}]} 6716            & 0.23   & 0.14   & 0.31        & 0.16  & 0.16   & 0.19    & 0.21    & 0.22    \\    
{[S\,{\sc ii}]} 6732            & 0.16   & 0.11   & 0.23        & 0.11  & 0.13   & 0.14    & 0.15    & 0.16    \\ 
{[S\,{\sc iii}]} 9069           & 0.40   & 0.33   & 0.25        & 0.51  & 0.40   & 0.40    & 0.40    & 0.34    \\  
{[O\,{\sc iii}]} 4363/5007 & 0.0030 &  0.0023 & $\leq$0.0041 &  0.034 &  0.0035 &  0.0026 & 0.0024 &  0.006 \\ 
{[S\,{\sc ii}]} 6717/6732       & 1.4    &  --    &   --        &  --   &  --    & 1.4     & 1.4     & 1.4     \\ 
$\eta'$                         & 0.3    &  --    &   --        &  --   &  --    & 0.27    & 0.18    & 1.3     \\ 
\end{tabular} 
\end{minipage} 
\newline
(1): line ratio measured in our scaned spectrum or the integrated spectrum 
of Ma\'\i z-Apell\'aniz (1999). \\
A, B, C and D are the line ratios in different parts of the nebula, 
from D\'\i az et al. (1987)
\end{table*}

\section{Modeling the H Balmer and He\,{\sevensize\bf i} absorption lines}

The optical continuum of an RH\,{\sc ii} is dominated by early-type stars. 
The spectra of O and B stars are characterized by strong H Balmer and He\,{\sc i} 
absorption lines, with very weak metallic lines formed in the photosphere of these 
stars (Walborn \& Fitzpatrick 1990). The H Balmer and He\,{\sc i} recombination 
nebular emission lines are superposed on the corresponding photospheric lines.
However, the higher order terms of the Balmer series and some of the He\,{\sc i} 
lines can be detected in absorption and the lower terms of the Balmer series can 
show absorption wings. The detection of these absorption features depends on the
spectral and spatial resolution of the observations, on the spatial distribution 
of the stellar cluster with respect to the nebular emission, and on the evolutionary 
state of the stellar cluster. These photospheric features have been detected in a 
spectrum of NGC 604 (Fig. 5) because it corresponds to the zone of 
the nebular hole where the core of the central cluster is located, so it maximizes
the contrast of the stellar with respect to the nebular contributions (see Fig. 2). 
In this section we will constrain the evolutionary state of NGC 604 and the IMF  using
the profile of the higher-order terms of the H Balmer series and the strength  of some
of the He\,{\sc i} lines in absorption.

\subsection{Description of the models}

Gonz\'alez Delgado et al. (1999b) have computed evolutionary stellar
population synthesis models that predict the photospheric absorption H Balmer and
He\,{\sc i} lines, between 3700 and 5000 \AA, for a single-metallicity stellar
population. The models, which are optimized for galaxies with active star  formation,
synthesize the profiles of the H Balmer series (H$\beta$, H$\gamma$,  H$\delta$, H8,
H9, H10, H11, H12  and H13) and the He\,{\sc i} absorption lines  (He\,{\sc i} 
$\lambda$4922, He\,{\sc i} $\lambda$4471, He\,{\sc i} $\lambda$4388, He\,{\sc i}
$\lambda$4144, He\,{\sc i} $\lambda$4121, He\,{\sc i} $\lambda$4026, He\,{\sc i} 
$\lambda$4009 and He\,{\sc i} $\lambda$3819), with a spectral sampling of 0.3 
\AA~pixel$^{-1}$, for a burst and for continuous star formation at a constant rate. 
They use a stellar library that includes NLTE absorption profiles for stars hotter 
than 25000 K, and LTE profiles for lower temperatures. The temperature and gravity 
coverage is $4000\leq T_{\rm eff}\leq50000$ K, and $0.0\leq log\,g\leq5.0$, respectively
(Gonz\'alez Delgado \& Leitherer 1999). The models assume that stars evolve 
from the main sequence following the evolutionary tracks of the Geneva group 
(Schaller et al. 1992; Schaerer et al. 1993a,b; Charbonnel et al. 1993; Meynet et al. 
1994). The strength of the Balmer and He\,{\sc i} lines is sensitive to the age after 
the first 3 Myr of evolution, and sensitive to the IMF if the age is younger than 
3--4 Myr. Models assume that stars have a metallicity $Z=0.008$, and make different 
assumptions about the IMF slope ($\alpha$=1.5, 2.35 and 3.0), and upper mass limit cut-off 
($M_{\rm up}=80$ M$\odot$ and 30 M$\odot$). For the purpose of this paper, the range of
ages computed spans from 0 to 10 Myr. 

\begin{figure}
\psfig{figure=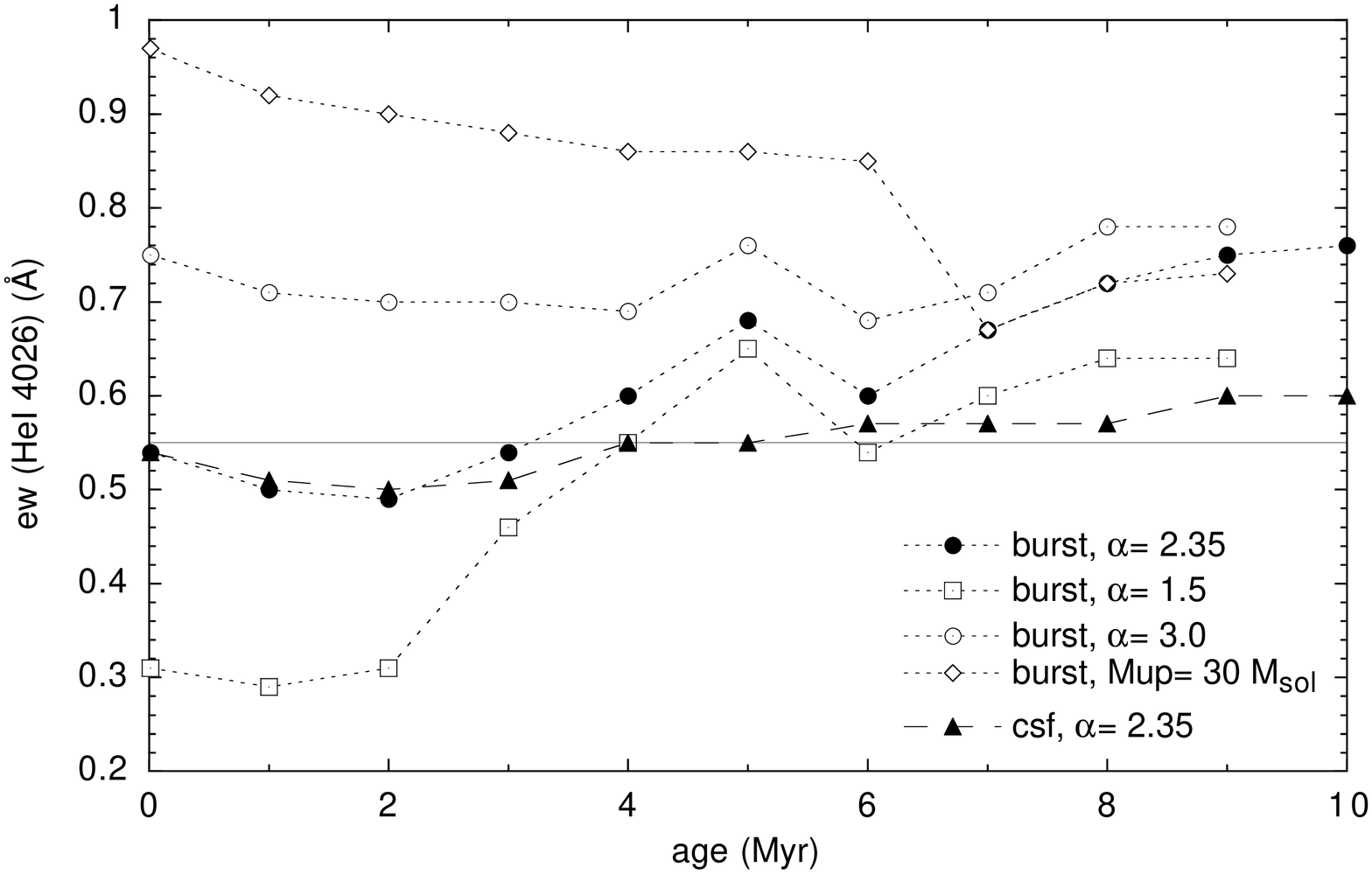,width=85mm,angle=0}
\caption{Predicted equivalent widths of the He\,{\sc i} $\lambda$4026 
stellar absorption line as a function of the age for continuous star 
formation (filled triangles) and burst (filled circles) models, assuming 
a Salpeter IMF between 1 and 80 M$\odot$. Burst models are computed 
for different IMF slopes, $\alpha$=1.5 (squares), 3.0 (open circles), 
and upper mass limit cut-offs, Mup=30 M$\odot$ (open diamonds). The observed 
value is plotted as an horizontal line. The equivalent widths of He\,{\sc i} 
$\lambda$4388, $\lambda$3819 are smaller than that of $\lambda$4026 and do not
add further constraints, so they are not shown for the sake of clarity.}
\label{ew} 
\end{figure}

\subsection{Model Results}

We use the equivalent width of He\,{\sc i} $\lambda$4388, $\lambda$4026, and 
$\lambda$3819, and the wing absorption profiles of H$\beta$, H$\gamma$, H$\delta$, 
H8, H9 and H10 to constrain the age, IMF and star formation law. He\,{\sc i} 
$\lambda$4388, $\lambda$4026, and $\lambda$3819 are detected in absorption, because 
the corresponding nebular emission lines have equivalent widths that are at least
a factor 3 weaker than the nebular emission in He\,{\sc i} $\lambda$4471 (Gonz\'alez
Delgado et al. 1999b). Note that He\,{\sc i} $\lambda$4471 is only partially filled
with the nebular emission; therefore, the nebular emission of the lines He\,{\sc i}
$\lambda$4388, He\,{\sc i} $\lambda$4026 and He\,{\sc i} $\lambda$3819 is $\simeq$0, 
and the equivalent widths of these absorption features represent well the strength of
the stellar continuum radiation of the cluster. The equivalent widths of these
lines, He\,{\sc i} $ \lambda$4388, He\,{\sc i} $\lambda$4026 and 
He\,{\sc i} $\lambda$3819, measured in the spectrum of Fig. 5 are 0.32, 
0.55 and 0.26 \AA, respectively. Their strength is compatible with continuous 
star formation and with burst models 3 Myr old (Fig. 10). However, He\,{\sc i} 
lines can constrain the IMF. Models with an IMF steeper than Salpeter or with 
$M_{\rm up}=30$ M$\odot$ predict He\,{\sc i} lines stronger than observed. The strengths 
of the lines are compatible with a 3 to 4 Myr old  burst formed following a Salpeter 
or slightly flatter IMF ($\alpha$= 1.5, Fig. 10). The absorption wings
of the Balmer lines are also compatible with burst models 3 Myr old,  with
Salpeter or slightly flatter IMF ($\alpha$= 1.5, Fig. 11).

\begin{figure}
\psfig{figure=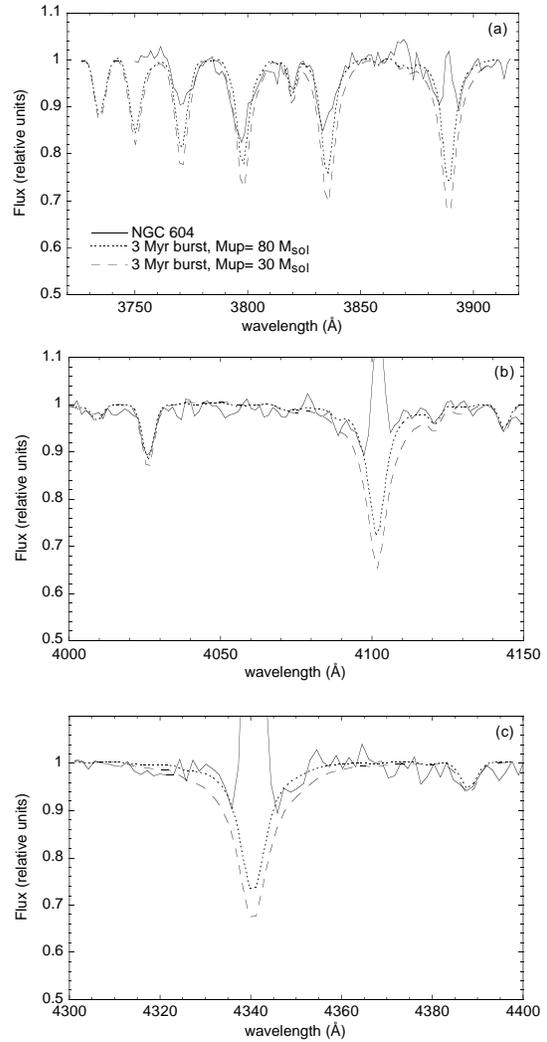,width=80mm,angle=0}
\caption{The normalized spectrum of Fig. \ref{optspec} (full line). 
Also plotted are the synthetic normalized spectra of a 3 Myr old instantaneous burst 
with a Salpeter IMF, and with $M_{\rm up}=80$ M$\odot$ (dotted line) and 30 M$\odot$ (dashed line). 
Models are degraded to the spectral resolution of the observations.} 
\label{optfit} 
\end{figure}

\section{Modelling the ultraviolet stellar lines}

The ultraviolet light from an RH\,{\sc ii} is dominated by O stars. These hot stars 
develop strong wind stellar lines due to the radiation pressure in ultraviolet 
resonance lines (Morton 1967). As a result, all the strong ultraviolet lines 
(e.g. O\,{\sc vi} $\lambda$1034, N\,{\sc v} $\lambda$1240, Si\,{\sc iv} $\lambda$1400, 
C\,{\sc iv} $\lambda$1550 and N\,{\sc iv} $\lambda$1720) show a blueshifted absorption 
(about 2000--3000 km s$^{-1}$) or a PCygni profile. The shape of the profile reflects 
the stellar mass-loss rate, which is related to the stellar luminosity, and thus to 
the stellar mass. Most of the ultraviolet spectra of RH\,{\sc ii} are dominated by 
absorption features (Rosa et al. 1984; Vacca et al. 1995; Mas-Hesse 
\& Kunth 1999), without any nebular emission, which is very similar to
those of starburst galaxies\footnote{One important exception is knot A in the 
RH\,{\sc ii} NGC 2363 (Drissen et al. 2000), so young that the effect of winds
is not yet reflected in its uv spectrum.}. Most
of these lines are formed in the stellar winds of the massive stars that belong  to the
starburst. The profile of these lines reflects the stellar massive content  in the
starburst; therefore, they depend on the IMF and star formation law (Leitherer, 
Robert \& Heckman 1995). In this section, we constrain the evolutionary state and the 
IMF of NGC 604 by means of fitting the profiles of the ultraviolet stellar lines
Si\,{\sc iv}  and C\,{\sc iv}.

\subsection{Description of the models}

Evolutionary stellar population models have been computed with the code 
Starburst 99 (Leitherer et al. 1999). The code uses a stellar library of 
{\it IUE} ultraviolet spectra of O, B and Wolf-Rayet stars (Robert, Leitherer 
\& Heckman 1993). The spectral resolution of the O and Wolf-Rayet stars 
observed is 0.75 \AA. These stars are located in the solar neighborhood. 
However, the evolutionary models computed here assume that the stars evolve 
from the main sequence following the evolutionary track at $Z=0.008$. Models 
are computed for instantaneous bursts between 0 and 10 Myr, and for continuous 
star formation lasting 10 Myr, and different assumptions about the slope 
($\alpha$=2.35, 3.0 and 1.5) and upper mass limit cut-off ($M_{\rm up}=$120, 80, 60 
and 40 M$\odot$) of the IMF. The model spectra are smoothed to the {\it IUE} 
spectral resolution of the observations of NGC 604, which is 6 \AA.

\subsection{Model results}

The strongest wind stellar features in the spectra of NGC 604 are N\,{\sc v} 
$\lambda$1240, Si\,{\sc iv} $\lambda$1400, C\,{\sc iv} $\lambda$1550, 
He\,{\sc ii} $\lambda$1640, and N\,{\sc iv} $\lambda$1720. These features 
are present in all the spectra of NGC 604; however, for this analysis we use 
the average spectrum of those at PA=149$\degr$, 136$\degr$ and 159$\degr$, 
and the spectrum at PA=32$\degr$, because they have the best 
signal to noise ratio. The evolutionary state, the star formation law and 
the IMF in NGC 604 is constrained with the profiles of these lines.

He\,{\sc ii} shows a broad emission profile in Wolf-Rayet and O3--O5 supergiant stars; 
N\,{\sc iv} shows a PCygni profile in Wolf-Rayet stars; N\,{\sc v} and C\,{\sc iv} 
show strong PCygni profiles in all O stars, and Si\,{\sc iv} only in O supergiants. 
Thus, the profile of these lines in the integrated spectrum of a starburst 
depends strongly on the stellar content and age of the stellar cluster. 
In particular, Si\,{\sc iv} shows a strong PCygni profile if the cluster 
formed in  an instantaneous burst and its age is between 3 and 5 Myr,
because within this age interval the ultraviolet light is dominated by O blue
supergiants. On the other hand, the profile of  C\,{\sc iv} depends
strongly on the IMF. It shows a strong PCygni profile if stars more  massive
that 60 M$\odot$ are formed in the cluster, and if the slope of the IMF is
flatter than $\alpha$=3.0. 

The spectrum of NGC 604 indicates that its stellar population must be 
dominated by massive, young O stars; thus, the stellar cluster must be young
and of short duration. Continuous star formation models can be ruled out
because they show Si\,{\sc iv} weaker than observed (Fig. 12). In
these models the line is diluted because the  fraction of O supergiants with
respect to the total number of O stars is lower in the continuous star
formation than in the instantaneous burst models. The strength of Si\,{\sc iv}
in NGC 604 indicates that it is an instantaneous burst. The age of the burst 
has to be between 3 to 5 Myr, because bursts younger than 3 Myr or older than 
5 Myr have very few O  supergiants, and thus very weak Si\,{\sc iv} (Fig. 13). 
On the other hand, from the strength of C\,{\sc iv} we can exclude instantaneous 
bursts with $M_{\rm up}\leq60$ M$\odot$ (Fig. 14), and IMF steeper than $\alpha$=3.0 
(Fig. 15). Therefore, we conclude that continuous star formation and instantaneous 
burst models with few very massive stars can be ruled out. The wind lines are 
compatible with an instantaneous burst, formed following a Salpeter or slightly 
flatter IMF, with upper mass limit cut-off higher than 60 M$\odot$. A 3 Myr 
instantaneous  burst with Salpeter and $M_{\rm up}\geq80$ M$\odot$ fits well 
the profile of the wind stellar lines.

\begin{figure*}
\psfig{figure=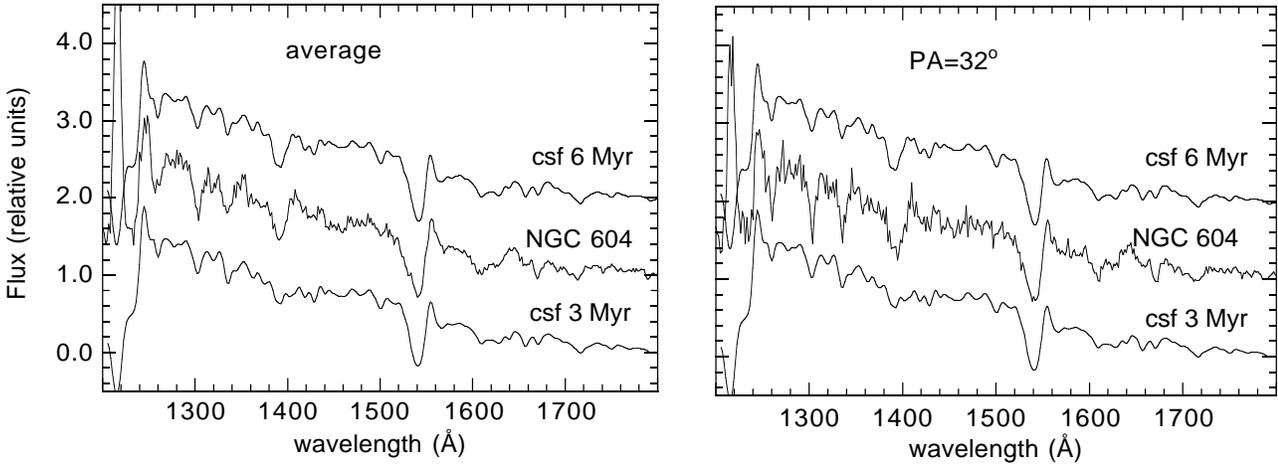,width=180mm,angle=0}
\caption{{\it IUE} spectra of NGC 604 (normalized at 1800 \AA), dereddened by 
a galactic extinction $E(B-V)=0.03$ using the MW extinction law, and by an
intrinsic extinction $E(B-V)=0.12$ using the LMC  extinction law, for the average
spectrum (a), and and intrinsic $E(B-V)=0.1$ for the spectrum  at PA=32$\degr$ (b).
Synthetic continuous star formation models for 3 and 6 Myr  are plotted in relative
units. The IMF is Salpeter with $M_{\rm up}=80$ M$\odot$.  Models are smoothed to the
resolution of the observations, 6 \AA.} 
\label{uvfit1} 
\end{figure*}

\begin{figure*}
\psfig{figure=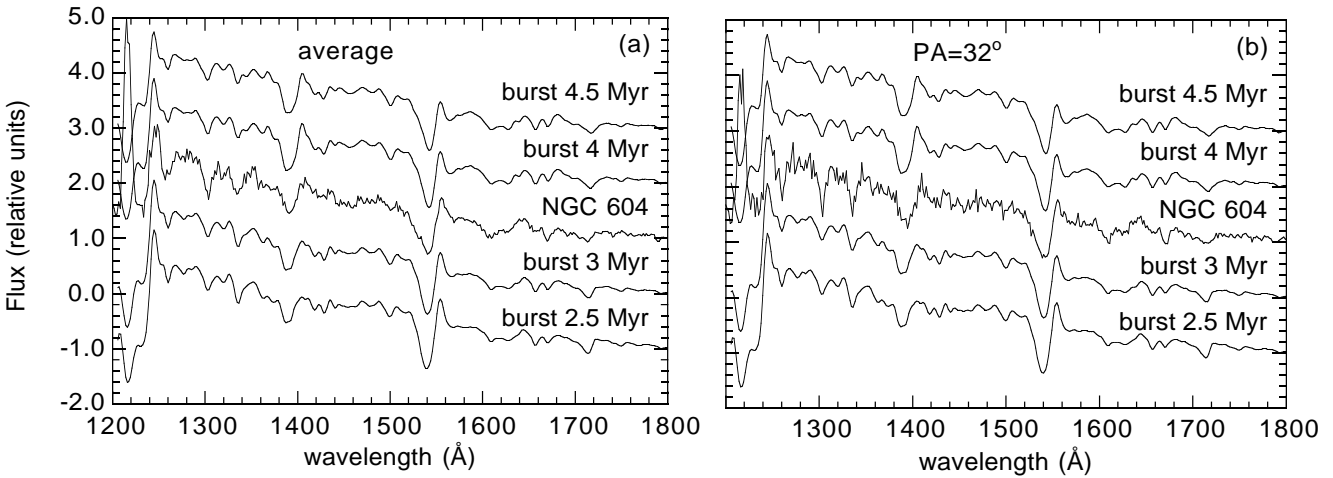,width=180mm,angle=0}
\caption{As in Fig. 12 for an instantaneous burst 2.5, 3, 4 and 4.5 Myr old.}
\label{uvfit2}
\end{figure*}

\begin{figure*}
\psfig{figure=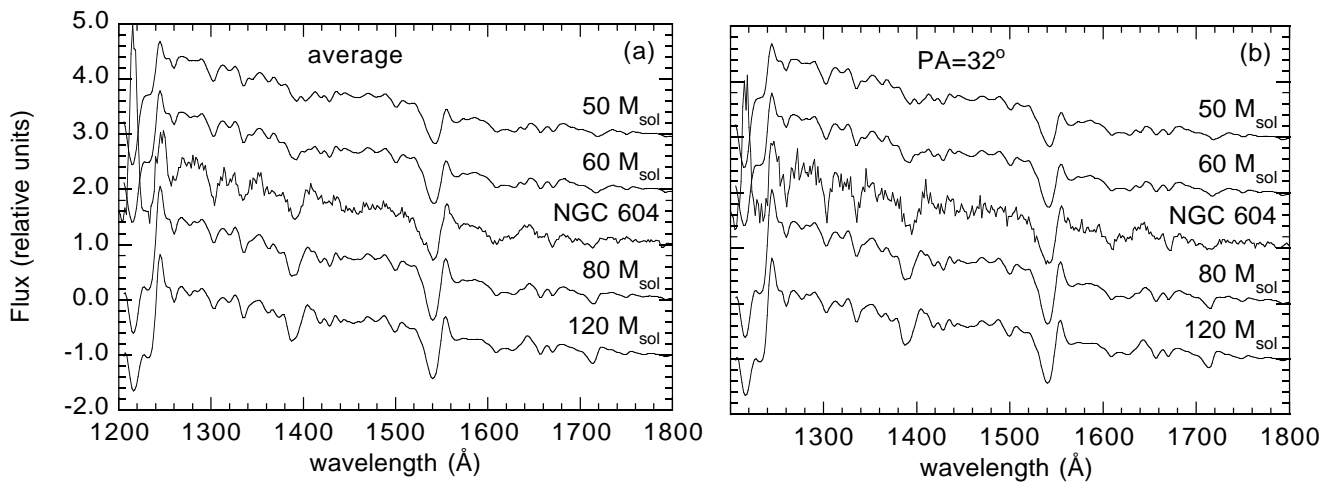,width=180mm,angle=0}
\caption{As Fig. 12 for an instantaneous burst 3 Myr old. The IMF is
Salpeter and  the upper mass limit cut-off as labelled.}
\label{uvfit3}
\end{figure*}

\begin{figure*}
\psfig{figure=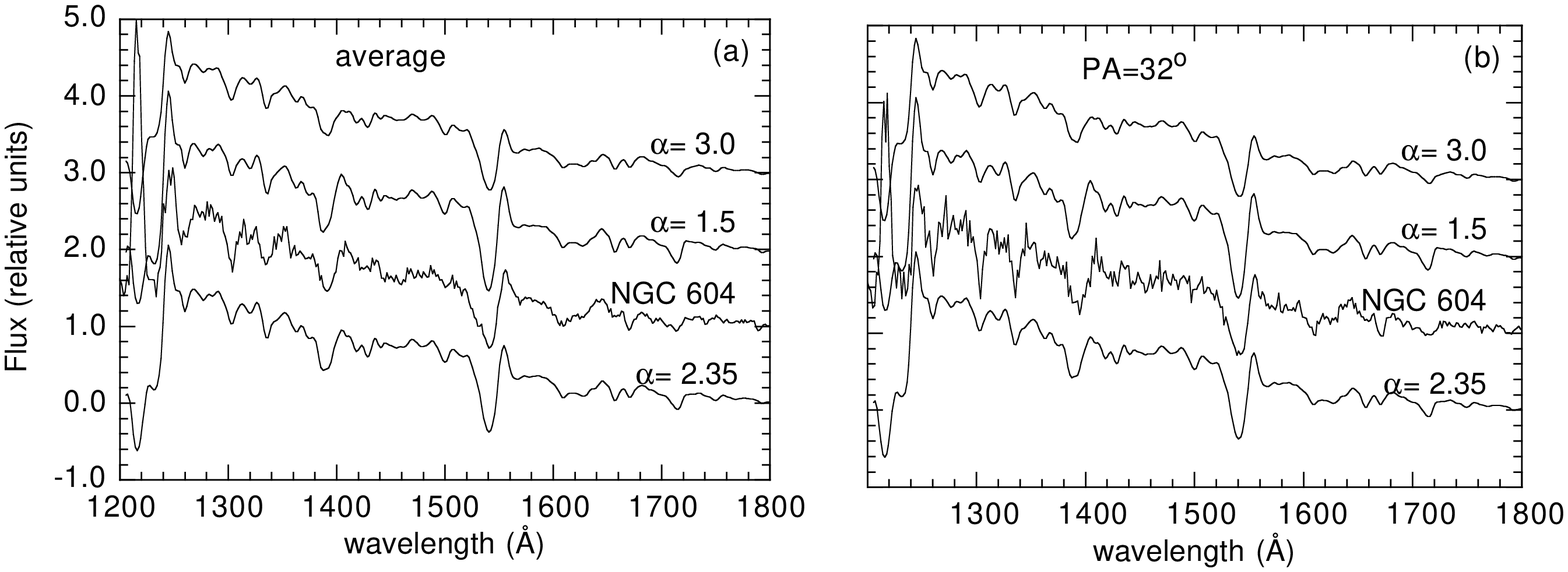,width=180mm,angle=0}
\caption{As Fig. 14. The IMF slope is $\alpha$= 2.35, 1.5 and 3.0, 
and $M_{\rm up}=80$ M$\odot$.}
\label{uvfit4}
\end{figure*}

\section{The massive stellar content of NGC 604}

We derive the massive stellar population in NGC 604 from the ultraviolet
continuum luminosity, comparing the observations with the predictions of 
synthetic models.  The number of Wolf-Rayet stars are derived from the
luminosity of the He\,{\sc ii}  $\lambda$1640 line. The stellar content
derived in this way is compared with that  derived from {\it HST} photometry
of stars in NGC 604. However, we need first to  estimate the extinction.

\subsection{Extinction estimates}

Leitherer \& Heckman (1995) have shown that the ultraviolet continuum 
arising from a young starburst ($\rm age\leq10~Myr$) has a spectral index, 
$\beta$ (F$_\lambda\propto\lambda^\beta$), which is independent of the IMF, 
star formation law and metallicity. Thus, any deviation from the predicted 
value, $\simeq2.5$ for a 3 Myr instantaneous burst, could be attributed to 
reddening.

The ultraviolet continuum flux distribution of the {\it IUE} spectra of NGC 604 
shows evidence of reddening, because after correcting by Galactic extinction, 
$E(B-V)=0.03$ (McClure \& Racine 1969), the spectra are flatter than the spectral
energy distribution  predicted by the evolutionary synthesis models. Massey \&
Hutchings (1983), analysing {\it IUE} spectra of H\,{\sc ii} regions in M33,
conclude that the M33 extinction curve is significantly different from that
of the Galaxy, because the spectra of the H\,{\sc ii}  regions show very weak
2200 \AA\ interstellar absorption dip, resembling those of the LMC and
SMC. Therefore, we use the LMC curve to match the observed spectra with the
spectral energy distribution of a 3 Myr old instantaneous burst. The $E(B-V)$
derived is 0.1 for the spectrum at PA=32$\degr$, and 0.12 for the average
spectrum of those at PA=136$\degr$, 159$\degr$ and 149$\degr$. These values are in
agreement with those derived by Massey \&  Hutchings (1983). However, the
extinction can be 0.03 higher than the values above if it is derived comparing 
the observed spectra with the spectral energy distribution predicted including
only the stellar contribution. This is due to the nebular contribution being
dominated by the two-photon continuum emission that peaks at $\sim$1500 \AA.
If the nebular continuum is not included, the spectral energy distribution is
steeper and a higher extinction is required to match the observed and
predicted ultraviolet flux distribution. The reason to do this new estimation
is because the nebula is more extended than the {\it IUE} aperture; thus, only a
fraction of the nebular continuum contributes to the {\it IUE} aperture. 

We have estimated also the extinction using the Calzetti, Kinney \& Storchi-Bergmann 
(1994) extinction law. This is an empirical extinction curve derived for 
starbursts, which is very similar to the MW extinction law but it does not 
show the 2200 \AA\ bump. Using this curve, we estimate $E(B-V)=0.2$ to match 
the observed spectra and the spectral energy distribution predicted by the models.

To distinguish between these two results, we compare the spectral energy 
distribution predicted by the evolutionary models with that of NGC 604. We build 
the spectral energy distribution of NGC 604 from the ultraviolet to near-infrared 
using the SWP+LWR {\it IUE} spectrum at PA=110$\degr$ plus that at PA=149$\degr$, 
and the average optical spectra scanning the inner 11$\times$60 arcsec 
(at PA=120$\degr$), and 14$\times$60 arcsec (at PA=60$\degr$). We use the $B$ 
(F439W) and $I$ (F814W) {\it HST}+WFPC2 images to flux calibrate the optical spectra, 
because the mode in which the scanning was performed precludes an accurate
absolute flux calibration based only on the ground-based observations. The 
resulting optical and ultraviolet spectra are normalized dividing by 
the flux at 2900 \AA. Then, the spectra are dereddened by $E(B-V)=0.1$ using 
the LMC extinction law, and by $E(B-V)=0.2$ using the Calzetti et al. (1994) 
extinction law. Even though apparently the spectra dereddened with Calzetti's 
law seem to match better the spectral energy distribution, a change in the scaling 
calibration factor of only 10 percent (larger than that used) gives a better 
match with the spectrum dereddened by the LMC law. Thus, given the uncertainties 
in the flux calibration of the ground-based optical spectra, we cannot really 
distinguish between these two extinction laws.

\begin{figure*}
\psfig{figure=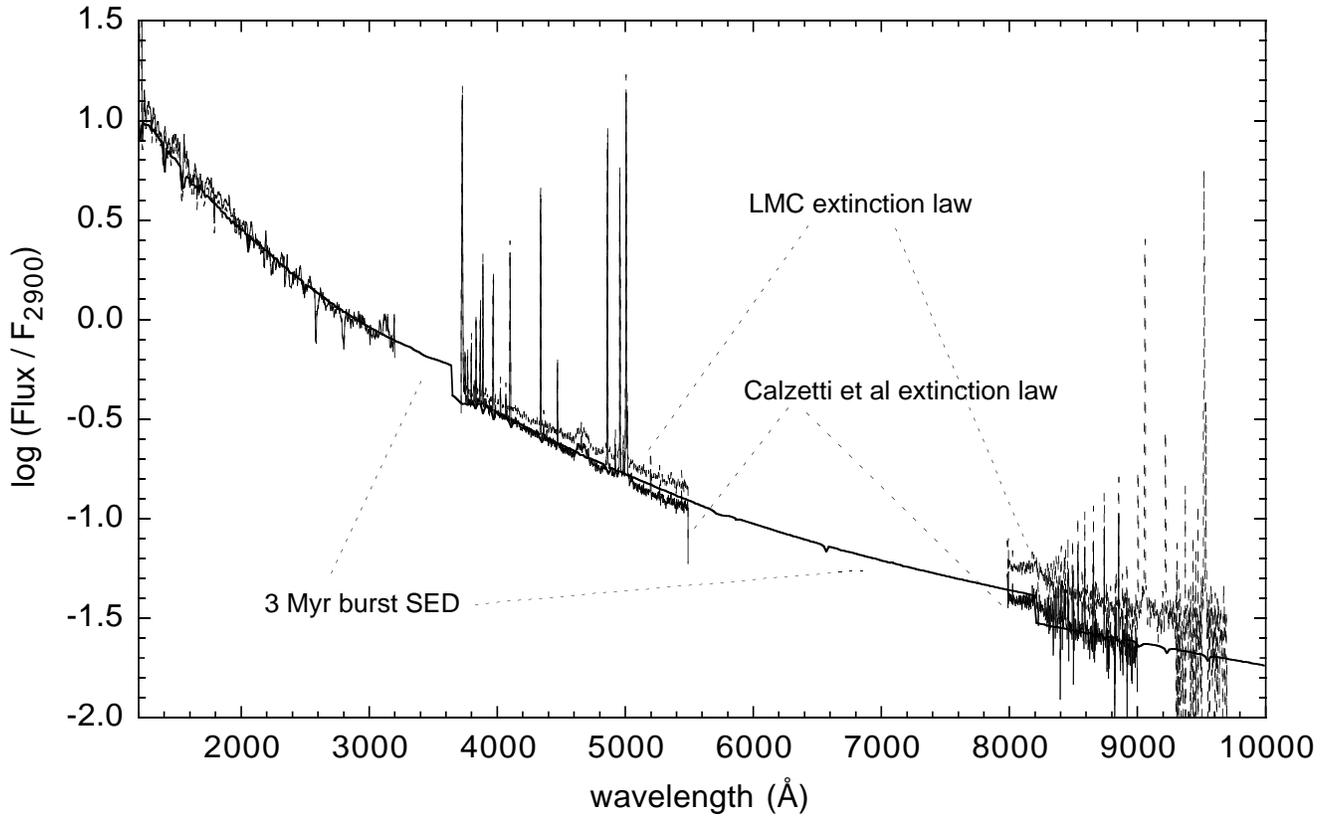,width=180mm,angle=0}
\caption{Spectral energy distribution of an instantaneous burst 3 Myr old, 
normalized at 2900 \AA, compared with ultraviolet plus optical spectra 
deereddned by $E(B-V)=0.1$ using the LMC extinction law, and by $E(B-V)=0.2$ 
using the Calzetti et al. (1994) extinction law.} 
\label{SED}
\end{figure*}

\subsection{The number of O stars and W/O ratio}

We assume that the total ultraviolet flux of the region is approximately 
the sum of the fluxes at PA=110$\degr$ and at PA=149$\degr$. After correcting 
by Galactic extinction, the flux at 1500 \AA\ is $5.1\times10^{-13}$ 
erg s$^{-1}$ cm$^{-2}$ \AA$^{-1}$. The intrinsic luminosity\footnote{Note 
that this luminosity and the quantities derived in Table 4 should be almost 
a factor two lower if the total ultraviolet flux is equal to the flux of only 
one of the {\it IUE} apertures, or to that measured in the F170W HST image, 
$L_{1700}\simeq 7\times10^{37}$ erg s$^{-1}$ \AA$^{-1}$.} is 10$^{38.16}$ 
erg s$^{-1}$ \AA$^{-1}$ (10$^{38.32}$ erg s$^{-1}$ \AA$^{-1}$) if the flux 
is corrected by $E(B-V)=0.12$ (0.2) using the LMC (Calzetti et al. 1994) 
extinction law.

The mass of the cluster, the ionizing photon luminosity and the number of O stars 
predicted by the evolutionary synthesis models are given in Table 4. These
quantities indicate that very massive stars must be present in the cluster
and that it is very young. In fact, instantaneous bursts older than 3.5 Myr
predict an ionizing photon luminosity which is at least a factor 3 lower than
the value derived from the H$\alpha$ flux, which is $log\,Q=51.54$ (s$^{-1}$). In
contrast, a $\simeq$3 Myr instantaneous burst  (with $M_{\rm up}\geq80\rm~M\odot$)
reproduces well the ionizing photon luminosity\footnote{However the values of $Q$
predicted are still a factor 1.5 lower than the photon luminosity derived from the
H$\alpha$ flux, suggesting that other sources in addition to the central cluster
contribute to the ionization of the gas, and/or that the extinction is higher than
$E(B-V)=0.1$.  Note that if log $L_{1500}=38.32$ ergs$^{-1}$ \AA$^{-1}$, then all the
quantities derived in Table 4 should be a factor $\simeq1.5$ larger.)}. These models
predict a mass of the stellar cluster that ranges between 0.1 and $2\times10^5$
M$\odot$, depending on the assumption about the upper mass limit cut-off and
slope of the IMF. The number of O stars ranges between 150 and 215 (if
$M_{\rm up}\geq80\rm~M\odot$), which is in agreement with the number of O stars
(186) reported by Hunter et al.(1996). 

The slope and upper mass limit cut-off of the IMF can also be constrained with
the observed Wolf-Rayet over O ratio, WR/O. {\it HST} images of the NGC 604 
cluster have detected 14 Wolf-Rayet or Of candidates (Drissen et al. 1993).
Thus, WR/O$\sim$0.075. This ratio is very high, suggesting that the
cluster is young (3--3.5 Myr), and that it must contain very massive stars.
Instantaneous burst models with the IMF slope steeper than Salpeter or
$M_{\rm up}\leq80$ M$\odot$ predict a ratio much lower than 0.075 (Fig. 17).
Models with IMF flatter than Salpeter or with upper mass limit
cut-off as high as 120 M$\odot$ reproduce well the observed value. 

\subsection{Deposition of kinetic energy in the interstellar medium}

Massive stars not only interact with the interstellar medium via their
radiation, but also depositing kinetic energy via their stellar winds. The
energy released during their lifetime is comparable to that deposited by a
supernova event (Leitherer, Robert \& Drissen 1992). 
The evolutionary state and the stellar content derived above for NGC 604
suggests that the massive stars in NGC 604 can release enough kinetic energy
to form many of the filaments and expanding shell structures observed in the
region. 

Yang et al. (1996) have reported the properties of five shells; these
have expansion velocities that range between 40 and 125 km s$^{-1}$, and sizes
between 35 and 125 pc (see also Sabalisck et al. 1995). The wind power required
to form these bubbles ranges between 2$\times10^{37}$ erg s$^{-1}$ and
9$\times10^{38}$ erg s$^{-1}$, with a sum total of $\simeq10^{39.25}$ erg s$^{-1}$. 

We estimate the wind power produced by the stellar cluster in NGC 604 using
Starburst99 (Table 4). For an instantaneous burst, the wind power ranges between
3$\times10^{37}$ erg s$^{-1}$ and 4$\times10^{38}$ erg s$^{-1}$, depending on the
assumptions about the slope and massive cut-off of the IMF. Thus, the energy
relased by winds in NGC 604 is not enough to explain the formation of the
five shells; however, it can explain the formation of the central hole (shell
number 3) and shell number 4 (see fig. 2 in Yang et al. 1996), that encircles
the core of NGC 604 cluster (cf. Fig. 2). The wind power required by
these two bubbles is 5$\times10^{37}$ erg s$^{-1}$ and 2$\times10^{37}$ erg s$^{-1}$, 
respectively.

\begin{figure}
\psfig{figure=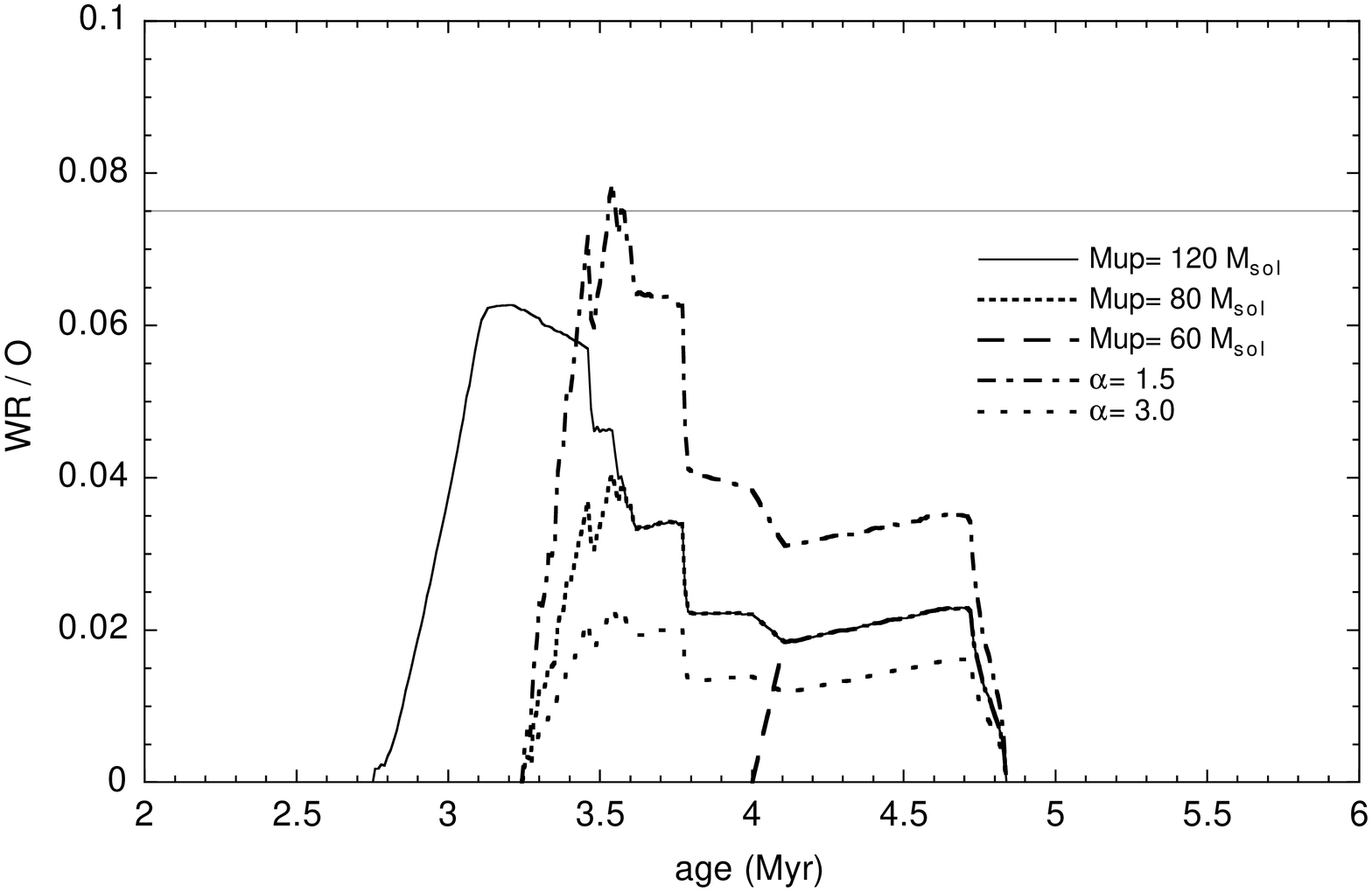,width=85mm,angle=0}
\caption{Ratio of the number of Wolf-Rayet with respect to the total O stars
as a  function of age, for different assumptions of the IMF slope, $\alpha$=
3.0, 2.35 and 1.5 ($M_{\rm up}=80$ M$\odot$), and upper mass limit cut-off, 
$M_{\rm up}=120$, 80 and 60 M$\odot$ ($\alpha$= 2.35).} 
\label{WRO}
\end{figure}

\begin{table}
\centering
\begin{minipage}{140mm}
\caption{Model predictions for massive stars and energy released.}
\label{tab4}
\begin{tabular}{@{}ccrcccc@{}}
age  & $\alpha$ & $M_{\rm up}$ & $log\,Q$ & Mass              & No.O & log
P$_w$\\  (Myr)&          & (M$\odot$) & (ph s$^{-1}$) & (10$^5$ M$\odot$) &  
   & (erg s$^{-1}$) \\
     &          &     & 
\multicolumn{4}{c}{$log\,L_{1500}=38.16$ erg s$^{-1}$ \AA$^{-1}$} \\
\\
3.0  &  2.35  &  80  &  51.28  &  0.47  &  209 & 37.98\\
3.5  &  2.35  &  80  &  51.20  &  0.43  &  189 & 38.59\\
4.0  &  2.35  &  80  &  51.10  &  0.51  &  218 & 38.38\\
4.5  &  2.35  &  80  &  51.02  &  0.62  &  244 & 38.50\\
5.0  &  2.35  &  80  &  50.86  &  0.72  &  262 & 37.36\\
\\
3.0  &  2.35  & 120  &  51.40  &  0.36  &  164 & 38.60\\
3.5  &  2.35  & 120  &  51.21  &  0.44  &  186 & 38.99\\
4.0  &  2.35  & 120  &  51.10  &  0.54  &  220 & 38.39\\
4.5  &  2.35  & 120  &  51.02  &  0.65  &  247 & 38.51\\
5.0  &  2.35  & 120  &  50.86  &  0.74  &  258 & 37.36\\
\\
3.0  &  1.50  &  80  &  51.28  &  0.13  &  180 & 38.05\\
3.5  &  1.50  &  80  &  51.23  &  0.12  &  152 & 38.76\\
4.0  &  1.50  &  80  &  51.11  &  0.17  &  199 & 38.55\\
4.5  &  1.50  &  80  &  51.06  &  0.24  &  239 & 38.67\\
5.0  &  1.50  &  80  &  50.85  &  0.31  &  267 & 37.41\\
\\
3.0  &  3.00  &  80  &  51.24  &  2.00  &  215 & 37.45\\
3.5  &  3.00  &  80  &  51.14  &  1.90  &  198 & 37.96\\
4.0  &  3.00  &  80  &  51.05  &  2.00  &  210 & 37.78\\
4.5  &  3.00  &  80  &  50.97  &  2.20  &  228 & 37.92\\
5.0  &  3.00  &  80  &  50.74  &  1.90  &  187 & 36.84\\
\end{tabular}
\end{minipage}
\end{table}

\section*{Summary and conclusions}

The main goal of this work is to constrain the evolutionary state of the giant
H\,{\sc ii} region NGC 604. For this purpose, we have analyzed the integrated
ultraviolet spectra taken by {\it IUE}, and optical ground-based spectra of the
region. The data are interpreted using evolutionary synthesis models optimized 
for star forming regions. These data are complemented with ultraviolet and 
H$\alpha$ images taken by {\it HST} with the WFPC2. 

The ultraviolet image shows that the ionizing cluster is spatially spread in
the inner 20$\times$20 arcsec, with the core of the cluster within a central
nebular hole shell structure. The optical spectrum, as it is well known, is 
dominated by nebular emission lines from the surrounding photoionized medium. 
In contrast, the ultraviolet spectrum of the region is dominated by 
absorption lines formed in the stellar winds of massive stars. However, other
photospheric stellar lines (the high order terms of the Balmer series and 
He\,{\sc i} lines) are detected at optical wavelengths near the Balmer jump. 
The spatial distribution of the stellar cluster with respect to the nebular emission,
allows to detect these lines in absorption in a spectrum corresponding to the 
inner 4 arcsec of the region, where the stellar light is maximized with respect 
to the nebular contribution.

The evolutionary state and the massive stellar content of the region is
derived in a self-consistent way using evolutionary synthesis and
photoionization models, applied to the ultraviolet resonance wind stellar lines, 
to the nebular emission-lines and to the higher-order terms of the Balmer
series and He\,{\sc i} lines in absorption. The three techniques applied suggest 
that the central ionizing cluster of NGC 604 is very young, 3 Myr old, with
no evidence for an age spread. The overall properties suggest that the
massive stars in the cluster were formed following a Salpeter or flatter IMF,
with presence of stars more massive than 80 M$\odot$.

Particular results from the modelling of the nebular emission lines include:

- The nebula is well described by a sphere of inner radius 20 pc and outer
radius 110 pc (determined by the ionization front). This value is
in agreement with the value derived from the surface brightness photometry of
the H$\alpha$ emission.

- The sphere is partially filled (filling factor $\simeq$0.1) with ionized
gas that is very tenuous, having electron density of $\simeq$30 cm$^{-3}$.

- [O\,{\sc i}]6300 and [O\,{\sc iii}]4363 emissions are well accounted for 
by photoionization.

Particular results from the modelling of the ultraviolet continuum are:

- The extinction affecting the stellar cluster is little, $E(B-V)=0.1$, if
the LMC extinction law is used to derive the value, or $E(B-V)=0.2$ if the 
Calzetti et al. (1994) extinction law is used. The 3 Myr instantaneous burst 
spectral energy distribution is well matched by the spectrum of NGC 604 
corrected with any of these two extinction laws. The value of the extinction 
is similar to the average extinction, $A_V=0.5$, derived for the gas by 
Churchwell \& Goss (1999).   

- The massive cluster provides a number of high energy photons that is enough
to photoionize the whole nebula. Thus, within the limits of the integrated
models that we develop for a uniform geometry, our results are compatible with
most of the ionizing radiation being reprocessed in the nebula, with no
significant escape of ionizing photons.

- The wind power provided by the massive stars of the cluster is enough to
form the central hole structure where the core stellar cluster is located.
However, it cannot provide all the wind power required for the formation of
the (at least) five shell structures seen in the H$\alpha$ image.

- The number of massive stars estimated is in agreement with that derived from
photometric studies based on the detection of individual stars.

\section*{Acknowledgments}

It is a pleasure to thank Gary Ferland and Claus Leitherer for kindly making 
their codes available, and Jaime Perea for the use of SIPL. Claus suggested 
that something should be done with those {\it IUE} spectra.
We have benefited from estimulating and helpful discussions with members of
the GEFE collaboration, in particular with  Angeles D\'\i az, 
Jes\'us Ma\'\i z-Apell\'aniz, Divakara Mayya, Guillermo Tenorio-Tagle, 
Elena Terlevich, Roberto Terlevich and Jos\'e M. V\'\i lchez. We are also 
indebted to Grazyna Stasi\'nska for his detailed comments from a thorough 
reading of the paper. The {\it IUE} helpdesk have always been solicituous 
with our queries, in particular we acknowledge Pedro Rodr\'\i guez Pascual 
and Rosario Gonz\'alez Riestra for support with the {\it IUE} spectra. 
{\it HST} images were retrieved from the ST-ECF {\it HST} archive.

\end{document}